\documentclass[sigconf]{acmart}
\AtBeginDocument{%
  }

\usepackage{xcolor}
\usepackage{multirow}
\usepackage{graphicx}
\usepackage{subcaption} 
\usepackage{array}

\newcommand{\edit}[1]{\textcolor{black}{#1}}

\usepackage{enumitem}

\begin{document}

\author{Alice Qian}
\email{aqzhang@andrew.cmu.edu}
\affiliation{%
  \institution{Carnegie Mellon University}
  \department{Human-Computer Interaction Institute}
    \city{Pittsburgh}
    \state{Pennsylvania}
  \country{USA}
}
\author{Ziqi Yang}
\email{ziqiyang@andrew.cmu.edu}
\affiliation{
    \institution{Carnegie Mellon University}
    \department{Human-Computer Interaction Institute}
    \city{Pittsburgh}
    \state{Pennsylvania}
    \country{USA}
}

\author{Ryland Shaw}
\email{rylandsh@usc.edu}
\authornote{Ryland Shaw was a pre-doctoral research assistant at Microsoft Research (Cambridge MA) at the time of this writing.}
\affiliation{%
  \institution{University of Southern California}
  \department{Annenberg School of Communication}
  \city{Los Angeles}
  \state{California}
  \country{USA}
}

\author{Jina Suh}
\email{jinsuh@microsoft.com}
\affiliation{%
  \institution{Microsoft Research}
  \city{Redmond}
  \state{Washington}
  \country{USA}}

\author{Laura Dabbish}
\email{dabbish@andrew.cmu.edu}
\affiliation{%
  \institution{Carnegie Mellon University}
  \department{Human-Computer Interaction Institute}
    \city{Pittsburgh}
    \state{Pennsylvania}
  \country{USA}}

\author{Hong Shen}
\affiliation{%
  \institution{Carnegie Mellon University}
  \department{Human-Computer Interaction Institute}
    \city{Pittsburgh}
    \state{Pennsylvania}
  \country{USA}}

\ccsdesc[500]{Human-centered computing~Interactive systems and tools}
\ccsdesc[500]{Computing methodologies~Artificial intelligence}
\ccsdesc[500]{Human-centered computing~Empirical studies in HCI}
\ccsdesc[500]{Human-centered computing~Empirical studies in collaborative and social computing}
\ccsdesc[500]{Social and professional topics~Computing / technology policy}

\renewcommand{\shortauthors}{Alice Qian, et al.}

\begin{abstract}
Responsible AI (RAI) content work, such as annotation, moderation, or red teaming for AI safety, often exposes crowd workers to potentially harmful content. While prior work has underscored the importance of communicating well-being risk to employed content moderators, designing effective disclosure mechanisms for crowd workers while balancing worker protection with the needs of task designers and platforms remains largely unexamined. To address this gap, we conducted individual co-design sessions with 15 task designers, 11 crowdworkers, and 3 platform representatives. We investigated task designer preferences for support in disclosing tasks, worker preferences for receiving risk disclosure warnings, and how platform representatives envision their role in shaping risk disclosure practices. We identify design tensions and map the sociotechnical tradeoffs that shape disclosure practices. We contribute design recommendations and feature concepts for risk disclosure mechanisms in the context of RAI content work. 
\end{abstract}


\keywords{AI safety, AI red teaming, risk, RAI content work, crowdsourcing, well-being, data work}

\title[Worker Discretion Advised]{Worker Discretion Advised: Co-designing Risk Disclosure in Crowdsourced Responsible AI (RAI) Content Work}

\maketitle

\section{Introduction}

\textbf{Responsible AI (RAI) content work} ~\cite{qian2025aura}, such as annotation, moderation, and red teaming of AI systems for safety, has become essential as organizations grapple with the risks of increasingly powerful AI systems. The rapid expansion of Generative AI (GenAI) has only heightened this reliance, fueling demand for large-scale human oversight to detect and mitigate safety concerns \cite{StanfordHAI2025AIIndex, GrandViewResearch2024GenerativeAI}. 
Meeting this demand requires workers to review harmful material, such as violence or manipulation, that requires nuanced human judgment~\cite{qian2025aura}. Such exposure can cause serious psychological distress, with studies documenting burnout, anxiety, depression, and in some cases post-traumatic stress disorder (PTSD) among workers~\cite{roberts2016commercial, qian2025aura, alemadi2024emotional, martinez2024secondary, spence2025content, gebrekidan2024content}.

Increasingly, organizations externalize RAI content work to \textbf{crowdsourcing platforms}~\cite{howe2006rise, berg2018digital}.  These platforms provide the infrastructure for on-demand, distributed labor markets, which AI companies then leverage to scale annotation, moderation, and adversarial testing \cite{udupa2023ethical, mann2025meta, egelman2014crowdsourcing}.
\textbf{Crowdworkers}~\cite{berg2018digital} may face significant \textbf{well-being risks} when taking on tasks that involve graphic violence, disturbing imagery, or to simulate harmful scenarios designed to probe AI system boundaries~\cite{qian2025locating}. 
Unlike RAI content workers employed directly by tech companies, who may receive institutional support such as mental health support or training ~\cite{qian2025aura, roberts2016commercial, steiger_psychological_2021}, crowdworkers often perform this work in isolation, without adequate preparation or access to support systems ~\cite{schlicher2021flexible, berastegui2021exposure}.


Prior HCI and CSCW work has documented structural conditions and power asymmetries in crowd work and proposed external tools like Turkopticon to increase transparency~\cite{gray2016crowd, silberman2018responsible, salehi2015we, irani2013turkopticon}. Yet platforms still provide few built in mechanisms to help workers anticipate or avoid emotionally harmful tasks. Existing content warning research focuses on social media audiences rather than paid workers in crowdsourced RAI tasks~\cite{prolific2025participant, prolific2025sensitive, ProlificAPIContentWarning2025, Zhang2024PerceptionsTriggerWarnings, shashirekha2023trigger, bell2025warning}. This gap motivates our focus on \textbf{risk disclosure}, the provision of upfront information about the nature and potential harms of content tasks~\cite{bharucha2023content, qian2025aura} as a \textbf{design challenge} embedded in sociotechnical systems that often obscure harm.
This design challenge is further complicated by tensions between stakeholder needs and incentives~\cite{finnerty2013keep, gaikwad2016boomerang, salehi2015we, irani2013turkopticon, salehi2018ink}. Prior work reveals that task designers\footnote{We use the term \textbf{task designers} rather than requesters to more accurately reflect the active and interpretive role individuals play in shaping the structure, framing, and content of crowdsourcing tasks. While requester is the platform-standard term, it implies a transactional relationship that downplays the design decisions and ethical considerations embedded in task creation. Following prior HCI work that emphasizes the creative and normative dimensions of task design~\cite{bragg2018sprout, qian2025locating}, we adopt task designer to foreground their agency and responsibility in shaping worker experience.} are pressured to recruit sufficient workers while still informing them of potential risks~\cite{qian2025locating, finnerty2013keep, kittur2008crowdsourcing, zheng2011task, bragg2018sprout} face pressure to recruit sufficient workers while also meeting ethical obligations to inform them of potential risks~\cite{qian2025locating, finnerty2013keep, kittur2008crowdsourcing, zheng2011task, bragg2018sprout}workers struggle for informed agency and fair compensation~\cite{irani2013turkopticon, salehi2018ink, martin2014being, silberman2018responsible, toxtli2021quantifying, schlicher2021flexible}, and platforms balance worker safety with efficiency and legal concerns~\cite{gillespie2010politics, xu2017incentivizing, xia2020privacy, allen2018design}. As Fieseler et al.~\cite{fieseler_unfairness_2019} remind us, crowdsourcing platforms mediate a ``triadic relationship'' between task designers, workers, and themselves, yet much research still treats this as a worker requester dyad; we therefore examine tensions across all three stakeholders in crowd based RAI content work and offer guidance for ethically grounded decisions that prioritize worker well-being while attending to the actors who shape task structure, disclosure policy, and enforcement.

To address this gap, we conducted one-on-one co-design sessions \cite{tang2024ai,kuo2023understanding} with 15 task designers, 11 workers, and 3 platform representatives to understand how risk disclosure decisions are made and what tools and workflows could better support effective disclosure practices. Our study focused on three key design dimensions: specificity (how detailed or granular the disclosure is), worker agency (the extent to which workers can act on disclosure), and task designer agency (the flexibility and support available to those creating disclosures). We ask the following research questions:\textbf{\textit{ RQ1:}} How do task designers, workers, and platform representatives perceive and approach different risk disclosure mechanisms across key design dimensions? and \textbf{\textit{RQ2: }}What tradeoffs and tensions emerge when stakeholders consider different configurations of risk disclosure mechanisms?

Our findings reveal that workers, task designers, and platforms hold divergent expectations about how risk should be communicated in RAI content work. Workers value clear, specific warnings and the ability to make informed decisions; task designers may fear that detailed disclosures will deter participation; and platforms aim to balance protection with non-intervention and scaling participation. These tensions manifest across key stages of task engagement, from sign-up to post-task feedback, exposing structural gaps in responsibility. We identify design opportunities to support disclosure practices such as adaptive filters and AI-assisted tools. Ultimately, we argue that risk disclosure is not just a technical feature but a sociotechnical negotiation of power, protection, and participation. 

Our contributions are threefold:
\begin{itemize}
\item \textbf{Empirical insights} into risk disclosure in crowdsourced RAI content work, based on \edit{individual} co-design sessions with 15 task designers, 11 workers, and 3 platform representatives. We surface how stakeholder roles, values, and constraints shape disclosure expectations and practices.
\item \textbf{\edit{An analysis of tensions among stakeholders}} of risk disclosure as a sociotechnical negotiation, identifying key tensions around specificity, worker agency, and task designer agency across the lifecycle of RAI content tasks. \edit{We surfaced these tensions by comparing perspectives across separately conducted co-design sessions with each stakeholder group.}
\item \textbf{Design implications} \edit{relevant to each stakeholder} for accountable and transparent risk disclosure systems.

\end{itemize}

\section{Related Work}

\subsection{Psychological and Well-being Risks of RAI Content Work} 
The rise of Responsible AI (RAI) initiatives draws new attention to the people behind AI systems: workers who annotate, moderate, and evaluate AI content. Their tasks -- collectively described as \textbf{Responsible AI (RAI) content work} \cite{qian2025aura}, including annotation, moderation, and red teaming for model safety -- are critical to system performance and societal impact. Yet their labor remains largely invisible or undervalued ~\cite{gray2019ghost}. 
A growing body of research documented the \textbf{psychological and well-being risks} workers face from viewing disturbing materials. Studies show that prolonged exposure to this material, coupled with limited support, contributes to burnout ~\cite{dosono2019moderation}, secondary trauma ~\cite{martinez2024secondary}, and PTSD ~\cite{steiger_psychological_2021, alemadi2024emotional, Michel2018ExContentMS, ruckenstein_re-humanizing_2020, Dwoskin_2019, arsht_2018_human}. Recent work reported that content moderators exhibit moderate to severe psychological distress and low well-being \cite{Spence2025ContentModeratorMentalHealth}. There are also broader harms, including privacy violations~\cite{pinchevski2023social, schopke-gonzalez_why_2022} and changes in workers’ personal beliefs and moral outlooks~\cite{newton_trauma_2019, Stackpole_2022, Douek_2021}. 

Although researchers are beginning to explore how organizations and individuals with decision-making power can support and address RAI content work psychological and well-being risks~\cite{qian2025aura, steiger_psychological_2021, bharucha2023content}, they focus on full-time \edit{content moderators with more formal employment contracts. Crowdworkers, by comparison, are less likely to have access to protection and support resources, like human resources divisions or institutional mental health resources, and have little recourse after facing emotional harm because of their lack of employment rights ~\cite{irani2013turkopticon, martin2014being, salehi2018ink, silberman2018responsible, toxtli2021quantifying, schlicher2021flexible}.} This is a significant gap, as crowdworkers increasingly perform similar high-risk annotation and moderation tasks, often with little to no warning about potentially graphic or disturbing content. However, far less is known about how these workers navigate such risks and how disclosure processes can be meaningfully designed to support them. 

\subsection{Risk Disclosure in Crowdsourced RAI Work: A Design Challenge}
While academic and organizational research emphasizes the importance of disclosing well-being risks in employed RAI content work settings (e.g., for internal content moderators~\cite{bharucha2023content, qian2025locating}), less is known about how such efforts are operationalized in crowdsourcing contexts. 

Most crowdsourcing platforms offer limited mechanisms for risk disclosure; however, some have recently made efforts to implement better disclosure practices. 
Prolific, for example, provides a researcher-facing ``sensitive content'' toggle during study design (see Appendix Figure \ref{fig:prolific_zoomed}) with an option to include additional information through keywords, along with guidelines to include warnings in task descriptions for both task designers~\cite{ProlificResearcherSensitive2025, ProlificAPIContentWarning2025} and workers~\cite{prolific2025participant}. Another approach is to limit exposure based on qualifications. Amazon Mechanical Turk (MTurk) provides task designers with an ``adult content qualification'' requirement option for workers age 18+  (see Appendix Figure \ref{fig:mturk_adult}). Research platforms such as Zooniverse and \edit{Toloka} provide examples of task-level warnings, but these remain inconsistent in format and enforcement.\footnote{https://www.zooniverse.org/}\footnote{https://toloka.ai/} However, the relative effectiveness of these methods is unclear, as well as how each stakeholder group involved in their use and implementation perceives them.  
Beyond disclosure options, platform-level governance varies significantly. 
Prolific allows workers to report tasks for review and potential removal, and anecdotal evidence suggests platforms like Toloka have occasionally removed tasks after worker complaints (see Appendix Figure \ref{fig:toloka_removed}). However, such interventions appear to be exceptions rather than rules.

Oversights in crowdsourcing platforms' disclosure practices are striking, given the effort put into building content warning systems for consumer-facing social media platforms. Platforms such as Tumblr, X, Instagram, and TikTok implemented content warnings to protect users from distressing material, ranging from tags like \#tw to overlays that blur graphic images. These systems aim to support informed consent and reduce exposure~\cite{Zhang2024PerceptionsTriggerWarnings, vit2025use}. However, their application is highly variable, and their effectiveness remains contested. Empirical studies show that while users generally appreciate warnings in principle, concerns persist about inconsistent enforcement, insufficient granularity, and unclear rationale~\cite{charles2022typology, bridgland2024meta, bell2025warning}. Some warnings may even increase anticipatory anxiety, especially among individuals with trauma experience ~\cite{sharevski2022meaningful}. These limitations point to the complexity of operationalizing harm reduction at scale, suggesting risk disclosure requires careful design tailored to user needs, platform affordances, and sociocultural context.

The context of crowdsourced RAI work differs in several ways. Unlike social media users, who encounter warnings during voluntary content consumption, crowdworkers engage with potentially harmful material as part of paid tasks. As such, risk disclosure in this domain intersects with distinct incentives such as compensation structures and data quality requirements that reshape the meaning of consent, exposure, and protection. These differences raise critical questions about how content warning strategies must be re-imagined for labor systems.
In this study, we examine how risk disclosure is approached, interpreted, and contested in this context. Accordingly, we frame risk disclosure as a complex design challenge that must account for divergent stakeholder needs, uneven power relations, and structural gaps in responsibility and support.

\subsection{Co-Designing Risk Disclosure Across Stakeholders}
\edit{While prior HCI and CSCW work has examined the structural conditions of crowdsourcing platforms and how responsibilities and harms are distributed across different actors, there is limited attention on how well-being risks are communicated, managed, or shared in the crowdsourced RAI content work space.}
Platforms mediate a ``triadic relationship'' between workers, task designers, and themselves~\cite{fieseler_unfairness_2019}, shaping interactions through design choices, algorithms, and governance practices. Foundational work like Turkopticon surfaced worker–requester power asymmetries~\cite{irani2013turkopticon}, and subsequent studies reveal trade-offs between efficiency and fairness embedded in task workflows, incentive systems, and reputation mechanisms~\cite{ho2015incentivizing, saito2019turkscanner}.

From the worker's perspective, platforms rarely center well-being. While crowdwork is often framed as flexible and empowering, that flexibility is often  illusory~\cite{rechkemmer2022understanding, varanasi2022feeling, liang2021embracing}. Workers face challenges with task rejection~\cite{mcinnis2016taking}, emotional strain~\cite{flores2020challenges, martin2014being}, and a lack of institutional protections~\cite{salehi2018ink, silberman2018responsible}. These harms are magnified in RAI content work, where workers are often exposed to harmful material without advance warning.
Task designers, meanwhile, focus on data quality, yet their choices, ranging from interface design to instruction clarity to compensation, significantly shape worker outcomes~\cite{wu2017confusing, han2020crowd, ho2015incentivizing}. Recent work calls for increased transparency and reflection among task designers, recognizing their role in structuring tasks that can cause or mitigate harm~\cite{zheng2011task, bragg2018sprout, qian2025locating, sutherland2018sharing}. Platform governance is another site to investigate how disclosure works in practice.
Platforms mediate access to tasks and enforce policies, yet both workers and task designers have limited insight into these mechanisms~\cite{toxtli2021quantifying, whiting2019fair, gray2016crowd}. 

\edit{In adjacent domains, researchers have begun to use explicitly multi-stakeholder approaches to navigate related issues. Work on content moderation has examined how platforms, moderators, and users negotiate responsibility for harmful material and participate in the design of reporting, flagging, and appeal systems, revealing tensions between safety, scale, and participation~\cite{jiang2023trade, glossop2025co, jhaver2018online}. Similarly, participatory design and co-design studies in gig and crowdsourcing contexts have brought together workers, clients, and platform representatives to surface conflicts around control, rating systems, and governance~\cite{huang2024design, hsieh_designing_2023}. These studies demonstrate the value of multi-stakeholder engagement in understanding how power and accountability are distributed, but they leave open how well-being risks are disclosed and managed in RAI content work; despite these advances, little is known about how stakeholders in this space actually perceive, interpret, and implement risk disclosure in practice.}
To address this gap, we conducted a multi-stakeholder investigation with task designers, crowdworkers, and platform representatives to understand how risk disclosure is approached in crowdsourced RAI tasks. In particular, we adopted a co design approach grounded in the belief that those most affected by system decisions should help shape them. Co-design foregrounds stakeholders’ lived experiences, surfaces tensions that may remain hidden in researcher-led or single-stakeholder processes, and in domains like gig work and freelance labor, helping redistribute agency in how problems are framed~\cite{huang2024design, hsieh_designing_2023}.
To our knowledge, this research is among the first attempts to apply co-design to examine risk disclosure in the crowdsourcing context, building on prior calls to design for multi-stakeholder participation~\cite{vines2013configuring} and framing disclosure not as simple transparency but as an ongoing negotiation of accountability, agency, and institutional power.

\section{Methods}

\begin{table*}
\fontsize{7pt}{10pt}\selectfont
\centering
\begin{minipage}[b]{0.32\textwidth} 
\centering
\begin{tabular}{lll}
\toprule
\textbf{PID} & \textbf{Job} & \textbf{Years} \\
\midrule
T1 & PhD Student & 3--5 \\
T2 & Data Analyst & 3--5 \\
T3 & Data Analyst & 1--3 \\
T4 & AI Researcher & 3--5 \\
T5 & Data Analyst & 1--3 \\
T6 & Data Analyst & 3--5 \\
T7 & Data Analyst & 3--5 \\
T8 & PhD Student & 3--5 \\
T9 & Research Assistant & 1--3 \\
T10 & Data Scientist & $>$5 \\
T11 & Data Analyst & 3--5 \\
T12 & AI Research Scientist & 1--3 \\
T13 & Data Analyst & 1--3 \\
T14 & AI Ethics Researcher & 3--5 \\
T15 & PhD Student & 1--3 \\
\bottomrule
\end{tabular}
\caption{Task Designers}
\label{tab:task-designer}
\end{minipage}
\hfill
\begin{minipage}[b]{0.32\textwidth}
\fontsize{7pt}{10pt}\selectfont
\centering
\begin{tabular}{lll}
\toprule
\textbf{PID} & \textbf{Job} & \textbf{Years} \\
\midrule
W1 & Crowdworker & 1--3 \\
W2 & Crowdworker & 1--3 \\
W3 & Crowdworker & $<$1 \\
W4 & Crowdworker & $<$1 \\
W5 & Crowdworker & 1--3 \\
W6 & Crowdworker & 1--3 \\
W7 & Crowdworker & 1--3 \\
W8 & Crowdworker & 1--3 \\
W9 & Crowdworker & 3--5 \\
W10 & Crowdworker & 3--5 \\
W11 & Crowdworker & $>$5 \\
\bottomrule
\end{tabular}
\caption{Workers}
\label{tab:workers}
\end{minipage}
\hfill
\begin{minipage}[b]{0.32\textwidth}
\fontsize{7pt}{10pt}\selectfont
\centering
\begin{tabular}{lll}
\toprule
\textbf{PID} & \textbf{Job} & \textbf{Years} \\
\midrule
P1 & Product Manager & 1--3 \\
P2 & Executive & 1--3 \\
P3 & Operations Associate & $<$1 \\
\bottomrule
\end{tabular}
\caption{Platform Representatives}
\end{minipage}
\end{table*}
\label{tab:platform-reps}

\subsection{Study Design}
We conducted \edit{individual} co-design sessions \edit{with different stakeholder groups} to explore how task designers, workers, and platform representatives perceive risk disclosure mechanisms \edit{and to explore the design space of how these mechanisms could be embedded into RAI content work}. Co-design was chosen as it foregrounds stakeholder perspectives and enables the surfacing of tensions through collaborative ideation. Drawing from participatory design traditions, we aimed to establish a ``third space'' where participants and researchers could meet on equal footing to engage in mutual learning and shared decision-making~\cite{muller_participatory_nodate, steen2013co, bodker_participatory_2018}. This approach aligns with prior HCI work that engages multiple stakeholders to understand underlying tensions when designing platform-level interventions~\cite{hsieh_designing_2023, tang2024ai}.

We opted to conduct individual co-design sessions with task designers, workers, and platform representatives rather than joint workshops involving multiple stakeholder groups. This decision was made for several reasons. First, separating stakeholder groups helped protect worker participants, many of whom are in structurally vulnerable positions. Prior research has shown that power asymmetries can shape who feels safe to speak and what perspectives are voiced in multi-party design engagements \cite{tang2024ai}. Conducting sessions separately allowed us to create safer and more comfortable spaces for open reflection, especially when discussing platform practices or task designer behaviors. Second, individual sessions allowed us to dive deeper into stakeholder-specific concerns, generating richer insights into risk perceptions, disclosure priorities, and contextual tensions that may not arise in broader group settings \cite{dillahunt2017reflections}. 
Finally, sessions with individual users allowed us to schedule sessions to accommodate participants' schedules, given the challenges of accessing all three stakeholder populations in this study.  \edit{Following previous studies \cite{kuo2023understanding}, when needed, we brought selected (anonymized) responses from previous sessions into later co-design sessions so that stakeholders could deliberate over others’ perspectives without direct confrontation, given known tensions in crowdwork power dynamics ~\cite{kittur2008crowdsourcing, irani2013turkopticon}. This design allowed us to surface tensions across stakeholders by juxtaposing how workers, task designers, and platforms each understood risk disclosure and where their priorities diverged.}

\subsubsection{\edit{Study Protocol}}
Each participant engaged in a single co-design session with one researcher, lasting approximately 60 minutes. \edit{We treated participants as design partners, inviting them to first propose new ideas, then to critique, modify, and extend the risk disclosure mechanisms we presented. For all three stakeholder groups, sessions followed a similar structure. First, we asked participants to reflect on existing challenges they faced when disclosing risk, viewing risk in tasks, and managing risk disclosure, and to walk us through at least one concrete example of a task they had designed, completed, or managed. We then invited them to describe what an ``ideal'' risk disclosure scenario might look like, probing for detailed accounts of hypothetical mechanisms and workflows.
To elicit deeper discussion, we prepared a series of design variations tailored to each participant group that varied along three design dimensions: (1) warning specificity, (2) worker agency, and (3) task designer agency (see Section 3.1.2 for detailed descriptions).}

\edit{Building on the task examples participants provided,\footnote{For participants who could not share task details due to company policy, we instead presented four example RAI content work tasks based on prior research~\cite{qian2025locating, qian2025aura}} we first walked all participants through different levels of \textbf{warning specificity} with Figma-based examples of how warnings might appear in task listings and interfaces.}
\edit{For task designers, we then concentrated our discussion on mechanisms that varied levels of \textbf{task designer agency}. We asked follow-up questions based on their earlier descriptions of how much control they wanted over risk disclosure, and showed design variations such as an AI feature that suggests a warning with an “accept and implement” option, AI-generated keyword suggestions for warnings, or automatic classification of tasks into sensitive content categories (e.g., hate speech).}
\edit{For workers, we explored possible features along the \textbf{worker agency} dimension, ranging from having no option to opt out, to binary options for opting into broad categories (e.g., ``sensitive tasks''), to more granular filters that excluded specific types of content (e.g., tasks involving ``cyberbullying and harassment'').}
\edit{For platform representatives, we discussed all three dimensions together, focusing on how different combinations of specificity, worker agency, and task designer agency might be implemented and governed at scale.} We used Figma 
to facilitate collaborative discussion. Sessions were audio/video recorded and transcribed for analysis.

\begin{table*}[t]
\fontsize{7pt}{10pt}\selectfont
\centering
\begin{tabular}{p{3cm} p{4cm} p{4cm} p{4cm}}
\toprule
\textbf{Dimension} & \textbf{Low} & \textbf{Medium} & \textbf{High} \\
\midrule
\textbf{Warning Specificity} & 
Sensitive vs. explicit or disturbing ~\cite{prolific2025sensitive, bell2025warning} & 
Category, modality, and keywords ~\cite{jhaver2023personalizing, bridgland2024meta} & 
Examples and severity scale ~\cite{jhaver2023personalizing}\\
\midrule
\textbf{Worker Agency} & 
No opt-out ~\cite{MTurkTutorial2017} &
Informed consent ~\cite{silberman2018responsible, ProlificParticipantSensitive2025} & 
Granular exposure control ~\cite{jhaver2023personalizing, jhaver2022designing} \\
\midrule

\textbf{Task Designer Agency} & 
Generic platform template ~\cite{prolific2025sensitive, ProlificAPIContentWarning2025} & 
Manual customization of warnings ~\cite{qian2025locating}& 
AI-assisted risk disclosure ~\cite{qian2025aura} \\
\bottomrule
\end{tabular}
\caption{Design dimensions for risk disclosure in crowdsourced RAI Tasks, organized by low, medium, and high levels of each dimension.}
\label{tab:design_dimensions_levels}
\end{table*}

\subsubsection{\edit{Risk Disclosure Dimensions}}

\edit{In this study, we focus on the following three key design dimensions: (1) \textbf{Warning specificity}: Prior work highlighted the amount of detail in content warnings as a recurrent point of friction:} studies of trigger and content warnings on social media show that warning systems range from simple binary alerts to detailed categories and examples, and that inconsistent implementation leads to confusion for both content creators and viewers~\cite{Zhang2024PerceptionsTriggerWarnings, bridgland2024meta}, \edit{and recent HCI research has drawn parallels between the practices of content moderation and emerging AI safety work, suggesting that similar dilemmas arise around how specifically to flag and communicate potentially harmful content to viewers and workers alike~\cite{qian2025aura}.}
\edit{(2) \textbf{Worker agency}: Workers sit at the center of crowdsourcing workflows yet have historically had limited control over how tasks reach them and how their concerns are heard in platform governance~\cite{irani2013turkopticon, toxtli2021quantifying}.}
To capture worker agency, we looked to scholarship on worker autonomy and worker‑driven advocacy. Projects like \emph{Ink} and \emph{We Are Dynamo} emphasize giving crowd workers more control over their work identities and opportunities~\cite{salehi2018ink,salehi2015we}. Research quantifying invisible labour and highlighting the burdens of algorithmic management similarly calls for greater transparency and opt‑out mechanisms~\cite{toxtli2021quantifying}. We mirror this push for agency in our continuum from no opt‑out to informed consent to granular exposure controls.
\edit{(3) \textbf{Task designer agency}: We aim to foreground task designers as active actors in shaping workers’ exposure to risk, and because there is comparatively less work that centers their constraints and ethical challenges relative to the robust literature on workers and platforms~\cite{gaikwad2016boomerang, papoutsaki2015crowdsourcing, kittur2008crowdsourcing, xia2020privacy}.} 
Much of prior research focused on novice task designers’ needs: tools like \emph{Fantasktic} and \emph{Sprout} help non‑expert requesters improve instructions and quality~\cite{gutheim2012fantasktic,bragg2018sprout}. Studies of novice researchers show that clear templates and guidance are essential when designing crowd tasks~\cite{papoutsaki2015crowdsourcing}. Emerging research identifying challenges of task designers being subject to organizational dynamics and possessing limited support in understanding potential harms to workers in the context of RAI content work further motivates the need to include this perspective~\cite{qian2025locating}. Our third dimension spans generic templates, manual customization, and AI‑assisted disclosure to reflect these varying levels of expertise. \edit{Following ~\cite{fieseler_unfairness_2019, gillespie2010politics}, we conceptualize platforms as mediating actors that configure the kinds of control available to workers and task designers, and that pursue their own goals around efficiency, scale, and liability through these controls. Therefore, platform agency in this study is treated as infrastructural and governance-level and is woven throughout the study, rather than as a separate design dimension.} 

\subsection{Participants}
We recruited 29 participants in total, including 11 crowdworkers, 15 task designers, and 3 platform representatives. \edit{For participants who identified as task designers and crowdworkers, we stopped recruitment when we reached thematic saturation \cite{caine2016local}. Recruitment for platform representatives was limited by access constraints \cite{caine2016local}}. Participants were recruited via LinkedIn and Reddit postings and referrals via snowball sampling \cite{parker2019snowball}. To recruit workers, we posted on crowdworker subreddits such as r\slash
mturk\footnote{https://www.reddit.com/r/mturk/} and also through the tasks requested by a few of our task designer participants who volunteered to assist with recruitment. \edit{Of the 26 task designers and workers, 5 were recruited via LinkedIn, 3 via Reddit, and 21 via snowball sampling. Platform representatives came from two different crowdsourcing platforms. }

We required that each group of participants have a base level of experience. For task designers, we ensured that all our participants had experience requesting at least one RAI content work task. We ensured all our participants who were workers had at least 1 month of experience completing RAI content work tasks and that platform representatives were currently employed at a crowdsourcing platform that hosted RAI content work tasks (see Table \ref{tab:inclusion-criteria}). In our selection, we asked task designer participants to indicate their background (i.e., job title and sector) and the types of RAI tasks they previously requested (e.g., prompt generation). Additionally, we asked prospective task designer participants to indicate what platforms they used for crowdsourcing as well as the types of content involved in their tasks (e.g., racism). Similarly, we asked workers to describe the types of RAI content work tasks they had experience completing, providing examples of each (e.g., ``provide a prompt for an AI model to produce violent content''). We also asked workers to estimate how many tasks they typically do in one day, the types of sensitive content they encounter, the platforms they use, and to provide an example of an RAI content work task they completed and a short description of how they felt after completing the task. \edit{As a result, we note variation in participants utilizing different platforms across task designers and workers (see Appendix Table \ref{tab:participant_summary_task_designer} and Appendix Table \ref{tab:participant_summary_worker} for aggregate information on task designer and worker participants)}. Tables \ref{tab:task-designer}-\ref{tab:platform-reps} summarize participant demographics. Our participants were based in the US, UK, and EU and were at least 18 years of age at the time of study participation.

\begin{table}[ht]
\fontsize{7pt}{10pt}\selectfont
\centering
\begin{tabular}{p{2cm} p{5cm}}
\hline
\textbf{Participant Group} & \textbf{Inclusion Criteria} \\
\hline
Task Designers & Must have experience requesting at least one task relevant to Responsible AI (RAI) content work. \\
\hline
Workers & Must have at least one month of experience completing RAI content work tasks on a crowdsourcing platform. \\
\hline
Platform Representatives & Must be currently employed or employed within the last year at a crowdsourcing platform that hosts RAI content work tasks. \\
\hline
\end{tabular}
\caption{Inclusion criteria for study participants across stakeholder groups.}
\label{tab:inclusion-criteria}
\end{table}


\subsection{Data Collection and Analysis}
All co-design sessions were conducted by the lead researcher, while analysis was performed collaboratively by the first two authors. Data sources included session transcripts and notes from co-design sessions. We applied the method of reflexive thematic analysis~\cite{clarke2017thematic, smith1995semi}. We conducted our qualitative analysis in multiple stages. First, two researchers independently generated fine-grained, initial codes for each transcript \edit{(e.g., \textit{``adding additional keywords may be a cognitive burden of extra content to read before starting a task''})}. \edit{These were then grouped into themes (e.g., ``expectations for worker protection'' to surface relationships across participant responses. Through iterative cycles of code consolidation, discussion, and memo-writing, we refined these into higher-level themes.} Discrepancies in interpretation were resolved through discussion to strengthen validity. Throughout, we used analytic memos and reflexive notes to document evolving insights and to support reflexivity across the analysis process.

This study was approved by our Institutional Review Board (IRB). All participants provided informed consent and were compensated \$30 USD\footnote{Participants in the UK and EU countries were compensated with the equivalent of \$30 USD} for their time via an online gift card. Given the sensitive nature of risk disclosure, we designed activities to minimize potential distress (e.g., rather than showing actual harmful text, we provide a placeholder: ``example of sentence featuring hatespeech''). Participants could withdraw at any time without penalty and were reminded not to provide information they were not comfortable speaking about. 

\subsection{\edit{Positionality Statement}}
\edit{This work was conducted by researchers based at North American research institutions, trained primarily in human–computer interaction, computer science, and communication. As academics whose careers benefit from studying crowdsourcing and AI safety, we recognize that we are structurally closer to task designers and platform staff than to the crowdworkers whose labor underpins these systems. To account for our positionality, we treated worker well-being as the primary normative anchor of our analysis: when accounts conflicted, we prioritized workers’ descriptions of harm and skepticism about protections, and read task designer and platform narratives as 
situated rather than neutral. Our focus on worker agency and task designer agency as design dimensions reflects both a commitment to expanding workers’ control over their exposure to risk and a recognition that task designers operate under organizational and platform constraints. We also recognize that emphasizing designable levers (e.g., warnings, filters
) can understate systemic issues in platform governance and labor regulation, so we frame proposed design interventions as complements to, rather than substitutes for, stronger structural protections for crowdworkers engaged in RAI content work.
}

\section{Findings}
\edit{Below, we present key findings from our analysis, organized across stages of RAI content work: Platform sign-up, platform preferences, decision to participate, task completion, and post-task completion. Each section follows a stage, with subsections covering design challenges (e.g., ``setting expectations for worker protection''). Within each subsection, we first examine how relevant groups of participants perceived and approached risk disclosure (RQ1), then surface the tradeoffs and tensions that arise across stakeholders (RQ2). Across all stages, participants reflected on current practices and proposed new designs such as filters, warning templates, and feedback workflows, exploring the design space of how risk disclosure should be embedded into RAI content work.}

{
\small
\begin{table*}[t]
\fontsize{7pt}{10pt}\selectfont
\centering
\begin{tabular}{|
  >{\centering\arraybackslash}p{0.3cm}|
  >{\raggedright\arraybackslash}p{1.5cm}|
  >{\raggedright\arraybackslash}p{5cm}|
  >{\raggedright\arraybackslash}p{2.5cm}|
  >{\raggedright\arraybackslash}p{3cm}|
  >{\raggedright\arraybackslash}p{2.5cm}|}
\hline
\textbf{\#} & \textbf{Task Stage} & \textbf{Design Challenge} & \textbf{Worker Priority} & \textbf{Task Designer Priority} & \textbf{Platform Priority}\\
\hline
4.1 &
\raggedright \textbf{Platform Sign-up}
&
Setting expectations for worker protection &
Platforms' protection from sensitive tasks &
Large and diverse worker population &
Liability protection \\
\hline
\multirow[t]{2}{0.3cm}{\centering 4.2} &
\multirow[t]{2}{1.5cm}{\raggedright \textbf{Platform Preferences}}
&
Defining boundaries for accepting tasks &
To have diverse options for tasks &
Large and diverse worker population &
Limit participation for worker protection \\
\cline{3-6}
& & Supporting worker self-awareness &
Self-awareness of triggers &
Trust workers know what they cannot do &
Scaffold worker self-awareness \\
\hline
\multirow[t]{3}{0.3cm}{\centering 4.3} &
\multirow[t]{3}{1.5cm}{\raggedright \textbf{Decision to Participate}}
&
Warning workers without deterring participation &
Information to decide to opt in or not to task &
Convince workers to do the task &
Worker and task designer retention \\
\cline{3-6}
& & Establishing definitions of sensitive content &
Clear definition of content &
Autonomy in describing risk &
Provide definition in compliance with regulations and policies \\
\cline{3-6}
& & Balancing task designer autonomy with agency &
Consequences for failed risk disclosure &
Agency to disclose risk &
Mechanism to resolve disputes \\
\hline
\multirow[t]{2}{0.3cm}{\centering 4.4} &
\multirow[t]{2}{1.5cm}{\raggedright \textbf{Task Completion}}
&
Balancing task completion with freedom to stop &
Stop task if uncomfortable &
Full worker task completion &
Prevent task abandonment at scale \\
\cline{3-6}
& & Fair pay for RAI content work &
More payment for difficulty and well-being harm &
Quality worker performance &
Optimize labor costs while retaining workers \\
\hline
\multirow[t]{2}{0.3cm}{\centering 4.5} &
\multirow[t]{2}{1.5cm}{\raggedright \textbf{Post-Task Completion}}
&
Encouraging quality feedback &
Choice to give feedback &
Quality feedback &
Large-scale quality feedback \\
\cline{3-6}
& & Mitigating harms from failed risk disclosure &
Response from task designers &
Make up for mistakes &
Maintain trust in platform \\
\hline
\end{tabular}
\caption{\edit{We organized the design challenges identified in our findings by stages of task completion, and highlighted how each challenge surfaced different priorities for workers, task designers, and platforms in the context of RAI content work. Each row corresponds to a stage-specific design challenge, while the columns summarize what each stakeholder group prioritized at that point in the task process. Taken together, the table illustrates how the same design decision can simultaneously serve worker protection, data and participation goals, and platform-level concerns about liability and scale, often pulling in tension with one another.}}
\end{table*}
\normalsize
}

\subsection{Platform Sign-up}

\edit{In the following section, we discuss how our participants perceive risk disclosure when workers are first signing up for crowdsourcing platforms (RQ1) and design tensions emerging in this stage (RQ2). Workers described wanting meaningful protection from harmful RAI tasks from the moment they create an account, while task designers emphasized having access to a large and diverse pool of potential participants, and platforms focused on limiting their own exposure to complaints and legal risk. Together, these priorities establish the backdrop for designing risk disclosure mechanisms.}
\subsubsection{Setting expectations for worker protection}
\begin{figure}[htbp]
    \centering
    \includegraphics[
        width=0.5\textwidth,
        height=0.5\textheight,
        keepaspectratio
    ]{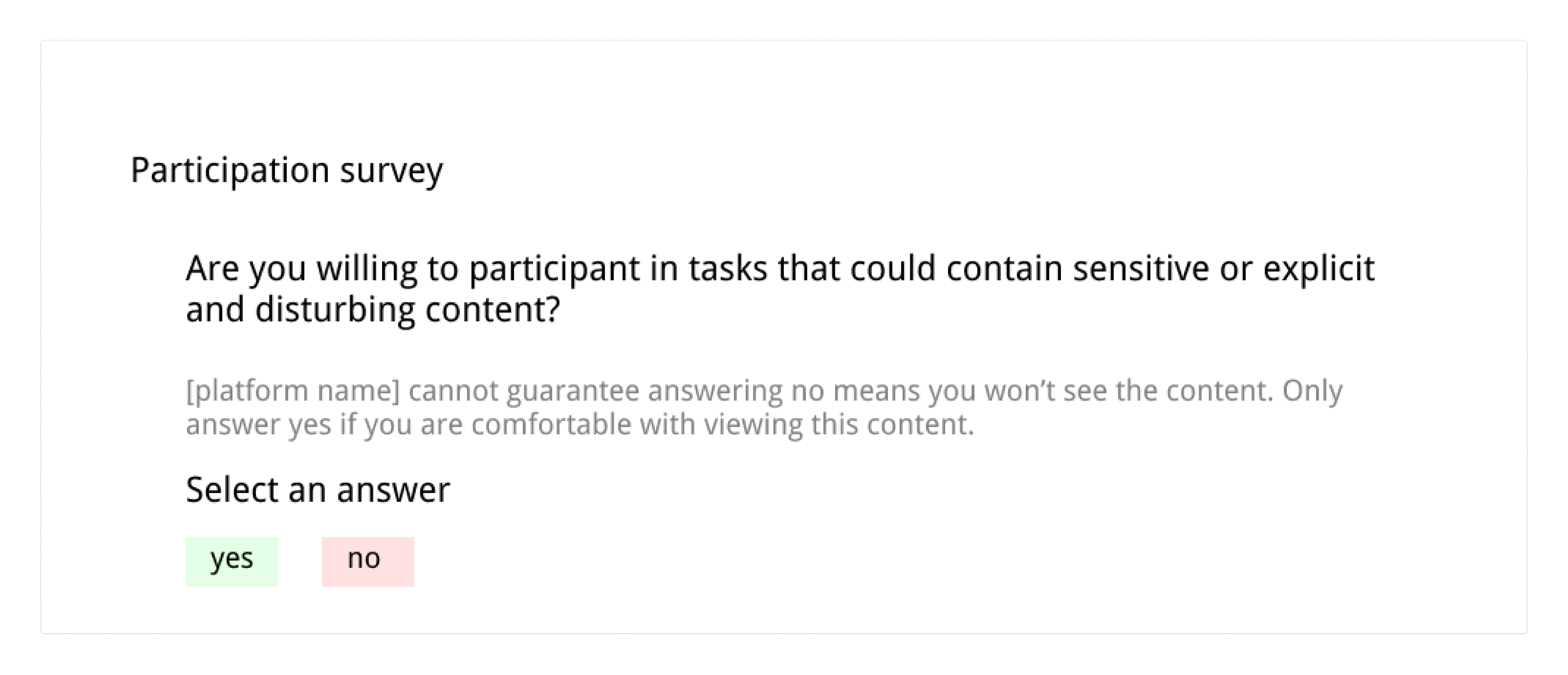}     
    \caption{\edit{Example of one possible participation survey interface used in co-design sessions. The example asks workers when signing up for a platform whether they are willing to participate in tasks that may contain sensitive or explicit content, while noting that the platform cannot fully guarantee shielding them from such material. Workers are given an option of responding with ``yes'' or ``no.''}}
    \label{fig:participation-survey} 
\end{figure}

\edit{Participants perceived opt-in and opt-out mechanisms as meaningful only if platforms actively uphold them.}
Although platforms sometimes provide such options for sensitive content, workers emphasized that platforms should be accountable for enforcing them, even if platforms may struggle to guarantee full shielding. When asked whether they were prompted about their tolerance for sensitive content upon first joining a platform, few could recall such questions. After discussing examples of how a platform might ask about willingness to view sensitive content \edit{by providing a ``yes-no'' option with a disclaimer about the possibility of still viewing potentially sensitive content} (see Figure \ref{fig:participation-survey}), workers expressed two sentiments. First, some, like W1, noted that platforms should actively uphold the preferences workers indicate: \textit{``I think the platform should protect [us]. [If a worker] wants to say no [and they] don't want to see [sensitive content]. It shouldn't be popping up anymore''} (W1). 

Others, such as W11, reflected that \textit{``from the frequency of the things I've been seeing I would say it's almost impossible for you to [not] be exposed to explicit contents \dots so I think it's okay [for the platform to be] vouching that you will not see [the content]''} (W11). However, W11 explained that if \edit{the design of the participation survey changed for the question to be more specific}--for example, asking about willingness to view racist content--it would be more important to enforce worker preferences, as they were personally harmed by that specific type of content (W11). From the platform perspective, P1 explained that such opt-in statements \textit{``gives the company or the employer some cover''}, but suggested that support the platform offers \textit{``could be more comprehensive. It could offer resources. You can say `if you experience issues, please reach out to us''} (P1). Moreover, P3 further explained that such options are \textit{``work best when paired with clear educational content''} (P3).
\edit{These perspectives reveal a tension between workers’ expectations of protection and platforms’ ability and willingness to guarantee it. Workers treated opt-in and opt-out settings as commitments that should shape their actual task exposure, whereas platform representatives highlighted practical challenges in ensuring that workers will never encounter sensitive content. As such, without additional support such as enforceable policies or reliable content filtering, opt-in mechanisms may risk serving as symbolic assurances.}
 
\subsection{Participation Preferences}
\edit{Our analysis of this stage focuses on how participants perceived risk disclosure when workers decided which tasks to see and accept on a platform (RQ1), and on the design tensions that emerged around participation rules and filters (RQ2). Workers sought control over which tasks they saw, balancing flexibility with income, while task designers prioritized large and diverse samples and platforms used filters and eligibility limits to gate sensitive tasks, revealing how participation controls simultaneously protect and restrict work and often shift risk management onto workers as task designers and platforms focus on coverage, diversity, and liability.}
\subsubsection{Defining boundaries for task participation}

\edit{Participants perceived participation decisions as an important way to manage risk, but they differed in how they thought those boundaries should be drawn.}
\edit{Some task designers saw platform-level restrictions as necessary to protect vulnerable workers.} For example, P1 described restricting self-harm tasks to an `expert' population, explaining that such individuals may be better equipped for exposure: \textit{``if someone works in counter terrorism, they're more readily [prepared]. I think they are a more prepared audience for dealing with the kind of content that might be [encountered]''} (P1). This desire to limit participation, however, conflicts with prior literature that emphasized the importance of diverse perspectives in responsible AI development~\cite{qian2025locating, dalal2024provocation}. This was further validated by several task designers in our participant pool (e.g., T8 and T7) who expressed the need for a diverse sample. 

\edit{Workers similarly viewed participation controls as both a protection and a constraint.} When we presented a range of options, from generic task filters to specific options for workers, participants expressed mixed preferences. Some workers echoed concerns that task filters could limit the number of tasks available and preferred instead to decide on a case-by-case basis using task warnings (W2, W3, and W9). Others favored category-specific filters (W5). On the other end of the spectrum, a few workers expressed they were not concerned about reduced task availability, stating \textit{``it's not really going to be much of an issue \dots as much as I care about the job, you know, I still have to take good care of myself. I don't want something that will probably affect [my] mental health in any way''} (W10). Ultimately, our findings surface a key tension: while restricting participation in tasks either by a platform or through worker preference can protect workers from exposure, it must also take into account workers' desires to view tasks they're interested in and task designers' needs to sustain participation. \edit{Taken together, both understood eligibility rules and filters as mechanisms for disclosure and protection, but they disagreed on how restrictive those mechanisms should be.}

\edit{These divergent views produced tensions around who gets to work, what counts as ``sensitive'' enough to filter, and how to keep studies viable.}
Some task designers strongly expressed a concern about participation in their tasks being limited. T3 describes this tension:
\begin{quote}
    \textit{``[From a] researcher's perspective, you [will] begin to see that if a lot of people are actually turning on the sensitive [content filter] so they don't see these tasks. Then you start to see a reduction in study participants, and at the end of the day, you might not be able to get enough [participants] \dots [From] the workers' perspective,  a lot of them would most likely turn [the filter] on.''} - T3
\end{quote} 
\edit{Others focused on the bluntness of broad labels.} For example, T14 observed that task filters for a categorization of ``sensitive content'' may be too broad: \textit{``[the task] might be sensitive, but not as sensitive as a worker might imagine it to be [by] filtering it out, they don't get to work on it''} (T14). 

\edit{To navigate these tensions, some task designers and platform representatives proposed more fine-grained and worker-controlled solutions.} T2 suggested that rather than a broad categorization, workers are given \textit{``a bit of control, like [with an option to indicate] words that you don't want to see''} (T2). P3 indicated that their platform was exploring the possibility of implementing filters by \textit{``pre-classified topic''} but noted that \textit{``the main challenge is building and maintaining large, accurately labeled datasets of tasks to make this possible. Filtering by broader topics is promising, as long as classification is reliable''} (P3). \edit{These proposals point toward designs that aim to protect workers from unwanted exposure while still preserving access to tasks and maintaining diverse, sufficient participation.}

\subsubsection{Supporting worker self-awareness}
\edit{Participants generally treated self-knowledge as a prerequisite for using task filters effectively, but they differed in how confident they were that workers had it.}
Many task designers assumed that most workers could reliably identify their boundaries. For instance, T15 reflected that if a worker completed RAI content work for a week, they would be familiar enough with the content to be able to indicate what upsets them. Many workers expressed a similar sentiment that they knew what content could upset them. For example, W3 explained that they have never liked seeing violence, while W6 stated they never want to do tasks involving sex exploitation. 

\edit{At the same time, our data revealed that self-awareness was often partial, unstable, or overshadowed by economic pressures.}
On the task designer side, T15 later expressed a concern that \textit{``doing this type of work can be a little numbing, and workers may not be paying the best attention to their own well-being''} (T15). This concern was justified as some  workers reflected on how they had become accustomed to viewing harmful content over time (e.g., T5, W10). As W10 explained, \textit{``right about now, I'm used to reviewing and working on this kind of task''} (W10). Moreover, some workers, like W6, had to view a task that displayed a violent altercation to realize they never wanted to complete tasks that showed someone getting injured. Other participants explained that, regardless of whether they were aware of what upset them, they felt some workers just didn't care to protect their well-being: \textit{``I know for a fact that lot of people don't really care about [what's] sensitive or not \dots a lot of [workers] say they don't really care if [a task is] sensitive or not. They just want to do it and collect their money''} (W3). 

\edit{These accounts suggest that relying solely on workers’ self-reported preferences is an incomplete strategy for protection.} Not all workers are fully aware of their limits before encountering harmful content, and some may downplay their own discomfort in order to secure income. T15 accordingly suggested that platforms could offer features that help workers be more \textit{``introspective''} about what they are willing to see. \edit{This raises a broader design question: if task filters and opt-in settings presume reliable worker self-knowledge, how might platforms instead scaffold reflection and recalibration over time, so that workers can use these tools more effectively to protect themselves?}

\subsection{Decision to Participate}
\edit{At this stage, participants draw initial boundaries around which tasks workers see and accept, after they have joined a platform but before they engage with specific RAI content work. We found that workers were trying to preserve meaningful choice over the kinds of tasks they take on without sacrificing access to income, while task designers sought large and diverse samples, and platforms experimented with filters and eligibility controls to manage exposure and liability. Our analysis focused on two linked challenges: how participation rules and filters define who is invited to participate in sensitive tasks, and how much they rely on workers’ own knowledge about what they can safely handle. Below, we discussed in each subsection how participants perceived risk disclosure (RQ1) and potential tensions (RQ2).}
\subsubsection{Warning workers without deterring participation}
\edit{Participants agreed that warnings are central to managing risk, but differed in how they weighed them against participation and data quality.}
Many task designers worried that disclosing risk would deter workers from participating, even as workers described warnings as crucial for making informed decisions. Several, such as T2, specifically feared that flagging sensitive content would dissuade workers from enrolling. In this way, task designers in our study echoed prior findings that prioritize obtaining task data and sustaining participation over other considerations~\cite{qian2025locating, finnerty2013keep, kittur2008crowdsourcing}. Some task designers felt an ethical obligation to include a basic warning despite perceived risks to the quality and speed of data collection (e.g., T12). Moreover, some task designers argued that \textit{``people who try to be responsible will not worry about [having enough worker participation] because you always need to make sure that you accurately represent the task before worrying about task completion''} (T15). T2 also reflected that responsibility for participation in RAI content work is: \textit{``for the platform to work on, because [we are] the customers [of] platform that [we're] using''} (T2). 

\edit{Workers, in turn, described warnings as one factor among several in their decision making, but often emphasized them as critical for informed choice.} Some noted that they prioritized other considerations, such as whether a task seemed \textit{``fun''} (W2) or paid well (W3). However, many also stressed that content warnings helped them decide what they could handle. W9 describes what they want to know about a task:
\begin{quote}
    \textit{``What I want to know about [a task] is [with the] images that will be shown, what they actually contain. [If] \dots it's a task like I feel I wouldn't be comfortable [with] \dots You know, in the end, I'm also a human being. As I'm training the model, I have to consider my own personal [preferences] also. So if it's something I feel I'll be able to work with, I'll proceed to work on it.''} -W9
\end{quote}
In fact, some workers described cases where inadequate risk disclosure caused significant harm to them. W6 recounted abandoning a task because they found the content was too \textit{``brutal''} to continue. 

\edit{These perspectives surfaced a core tension: task designers feared detailed warnings would suppress participation even as workers saw them as essential for protection, and platform representatives offered initial design ideas to navigate this gap.}
P2 suggested that platforms tell workers \textit{``your likelihood of seeing [a type of harmful content] is 1 in 100 \dots Putting numbers, averages, and percent likelihoods behind [warnings] would be helpful \dots [and] would help [platforms] get more people doing the work''} (P2). P2 proposed a tradeoff: 
\begin{quote}
    \textit{``If you answer this harmful question, you get paid more, but if you decide not to, we will give you five extra non-harmful questions so that [workers] have the option of: `do you want to expose yourself to harm and get paid the same and do less work or get paid the same and do more work?'''} - P2
\end{quote}
However, these suggestions presuppose that platforms can supply enough non-RAI content tasks \edit{to make such options viable}. Given worker observations that RAI content work is increasingly prevalent (e.g., W11), it is unclear how feasible such designs would be in practice. \edit{Together, these findings suggest that the design of risk disclosure cannot be reduced to a question of messaging alone, but rather depends on broader platform-level choices about task offerings, incentives, and the willingness of organizations to engage in greater transparency with workers.}

\subsubsection{\edit{Establishing definitions of sensitive content}}
\edit{Participants across roles agreed that ``sensitive content'' should play a central role in risk disclosure, but differed greatly in how they defined it.}
Some task designers assumed that their definition of risky, sensitive, or explicit content was aligned with others' to the point they did not need to see how the platforms they used defined it. For example, T15 \textit{``very rarely''} read the platform's definition of sensitive content because they \textit{``intuitively know the type of content that\'s in there''} and the \textit{``platform typically would put [a] blanket statement \dots [that's] usually not useful to read''} (T15). However, when prompted to explain how they personally defined `sensitive' versus `explicit or disturbing' content, the explanations that several participants provided varied. Some participants interpreted sensitive topics as \textit{private information [about workers]''} (T1), while others understood them as topics that may trigger reactions in some workers but remain neutral for others, for example T11 noting that \textit{cyber bullying is sensitive because it triggers [you] if you've been cyberbullied before''} (T11). Moreover, T5 defined sensitive content based on examples of \textbf{viewer discretion warnings} they saw on television: \textit{``[in] movies on Netflix \dots there's always a warning that says `nudity or violence and drugs'''} (T5). These findings echo longstanding debates about how many decisions in content moderation are left to \textit{`I know it when I see it'} judgments~\cite{gillespie2020expanding, ohioknow}.

\edit{Workers revealed how such disagreements on definitions could undermine the effectiveness of warnings.}
W1 encountered a task that they found upsetting, even though it had not been framed as sensitive: \textit{``maybe to [the task designer who] posted it \dots there is nothing that sensitive. But [for the worker] doing it, I see it as a sensitive issue that [the worker] doesn't want to see''} (W1). W5 emphasized that a clear definition of task severity can signal when a task \textit{``is going to be deeply harmful or offensive''} (W5). Platform representatives, in turn, reported drawing on commonly known taxonomies of harm (P1) and existing regulations and policies (e.g., European Union's Digital Services Act~\cite{EU-DSA-2022}) to structure their categories. Yet they also stressed persistent challenges in \textit{``standardizing language (especially around `type' and `severity') and making severity ratings more objective,''} acknowledging that such definitions \textit{``can be highly subjective and vary significantly across cultures''} (P3). \edit{As a result, task designers, workers, and platforms often operate with overlapping but non-identical notions of sensitivity, leaving substantial room for mismatch. }


\edit{These mismatches fed into a broader tension between task designers’ desire for agency over how they disclose risk and workers’ calls for stronger accountability mechanisms.}
Several task designers indicated that they wanted agency to make the final decisions about how to disclose risk in their tasks. As T1 explained: \textit{``I know everything about my research, and I want to explain my research by my word[s] and by myself''} (T1). \edit{At the same time, many were open to receiving support in determining what warnings were most appropriate for their tasks.}
When prompted with scenarios involving  AI or prediction-based suggestions, some expressed optimism about tools that could recommend warnings based on how similar tasks had been labeled by other task designers (e.g., T3 and T12) or on what workers indicated in the past (e.g., T2). Others emphasized that quantitative recommendations would be particularly persuasive; T3, for example, explained that if a system indicated that \textit{``about nine times out of ten''} similar tasks included a warning, they would be inclined to follow that norm.

\edit{Platform representatives described similar explorations.} P3's platform explored AI-supported options to perform categorization: 
\begin{quote}
    \textit{``[The] mixed content [of RAI content work] makes accurate categorization difficult \dots extensive testing would be required before implementation \dots The primary barrier to doing this is the pressure of strict production deadlines and the fast pace of the industry, which limits time available for a thorough [review].''} - P3
\end{quote} 
\edit{This suggests an appetite for tools that could help reconcile divergent definitions of sensitivity, but also highlight substantial technical and organizational barriers.}

\edit{Even when task designers welcomed support, many emphasized that suggestions should guide rather than override their judgment, and that trust in how suggestions are produced is critical.}
Participants proposed design options, including having multiple suggestions to choose from (T2 and T3), an option to edit warnings rather than just accept suggestions (T7 and T11), and integrating AI options into specific questions about risk disclosure, such as suggesting keywords for a warning (T7). \edit{At the same time, T1 and T15 cautioned that a certain amount of trust in how the suggestion was formed--particularly if it was AI-based--was necessary for them.}
For example, T15 argued that suggestions based on how other task designers have disclosed risk would not be helpful unless the suggestion can clearly explain how similarity between tasks was calculated (T15).

\edit{From workers’ perspectives, accountability was paramount when definitions and warnings failed.} W7 argued that task designers \textit{``should be held responsible for anything that comes up''} in relation to their content warnings and also felt platforms should enforce consequences for inadequate risk disclosure because it is their job \textit{``to make their platform more [worker] friendly [so] everybody will be satisfied''} (W7). Some task designers expressed sentiments in agreement that platforms should play an active role in holding other task designers accountable (e.g., T2, T8, and T9). However, tensions arose regarding how such mechanisms might be implemented without undermining task designers' autonomy. W3 suggested a platform-level review after disclosure of information was entered, whereas T1 expressed discomfort with being \textit{``scanned or judged''} regardless of the accuracy of an AI review (T1). While participants like T15 argued that task designers genuinely concerned about worker well-being would not be troubled by platform oversight, T1's concern illustrates a key risk: task designers who feel over-policed or misjudged may simply move their work to another platform. \edit{Overall, our findings highlight a delicate balance: aligning definitions of sensitive content and strengthening accountability requires tools and governance that support task designers while still centering workers’ needs.}

\subsection{Task Completion}
\edit{Task completion captured what happened once workers had committed to an RAI content task: workers tried to preserve ongoing consent and protect their well-being while earning income, task designers prioritized full completion and high-quality outputs, and platforms sought to minimize abandonment and control labor costs. Our analysis focused on two interlinked challenges: how workers can exit or skip harmful tasks without penalty, and how pay should reflect the emotional and cognitive burdens associated with this work. These dynamics demonstrated how design choices regarding stopping rules and compensation influence workers’ ability to protect themselves once a task is underway. We discussed in each subsection how participants perceived risk disclosure (RQ1) and potential tensions (RQ2).} 
\subsubsection{Balancing task completion with freedom to stop}
\edit{Workers’ accounts highlight a tension between the need for ongoing consent and the pressure to complete tasks.}
Even after opting into a task, several described encountering unexpected harms or discomfort.
W2, for example, recounted a task that became increasingly distressing: \textit{``I found some section of the task way too sensitive, which actually stressed me [out], but I had no option than to complete it. I wish there was that flexibility, that I could skip those parts''} (W2). Such sentiment highlights how workers wanted the ability to stop or skip harmful segments, but felt locked in by task and platform expectations. Exercising this freedom to stop was complicated by concerns about pay and reputational penalties, with several participants arguing that leaving a task due to its content should still warrant at least partial payment for the work already completed (e.g., W4–W6). As W3 explained: \textit{``if [completing the task] has a very high effect on [my] mental health, I don't really care if I'm paid or not. If I'm not comfortable with doing I'm not doing it, but it shouldn't have a penalty.''}

\edit{Both workers and some task designers pointed to partial payment models as a promising way to reconcile these concerns.} W4 provided a suggestion for platform enforcement:
\begin{quote}
    \textit{``[The] platform should inform [task designers] to to pay some amount of money, since it was stated that you're going to be getting [a specific] amount after completion \dots because what [the task designer] say[s is] sensitive might not be actually what, what [I] perceive---like sensitive might be extreme to me, like very sensitive. So they should be able to compensate for that.''} - W4
\end{quote}
Some task designers endorsed this approach. For example, T15 noted that they would still want to use data from workers who did not finish a task and that partial compensation was reasonable in these cases. \edit{As such, we find that platform-supported partial payment may help align ongoing consent with economic security, allowing workers to exit harmful tasks without bearing the cost.}

\subsubsection{Fair pay for RAI content work}
\edit{Participants viewed compensation for RAI content work as a central but contentious issue, with workers calling for higher pay and task designers and platforms weighing budget constraints and concerns about coercion.}
Echoing longstanding calls for fair payment \edit{in crowdwork~\cite{silberman2018responsible, irani2013turkopticon}}, several workers expressed a desire for increased pay specifically for RAI content work tasks. 
\edit{Workers described current rates and what they felt would be more appropriate for the burdens involved.} Some reported doing RAI content work at pay levels close to the standard U.S. minimum wage (for example, \$12.50 USD an hour for W11) and suggested higher rates, with proposing pay from \$30-\$50 USD (W1-W3 and W6). 

A number of task designers were, at least in principle, open to paying more for this type of work.
T14 proposed that platforms should charge 2-5\% more for such tasks, a margin that other task designers, like T4, regarded as a negligible increase. Others felt that any premium should depend on an assessment of whether a task's content was `severe' enough to warrant higher pay (e.g., T9). For some, paying more was also linked to accessing a more experienced pool of workers, such as platform designated `AI taskers' who had a proven track record with similar tasks and were presumed to be better able to handle exposure (e.g., T7 and T14)\footnote{https://participant-help.prolific.com/en/article/5baf0c}. P3 provided platform rationale for recommending higher pay: \textit{``the cognitive load of repeatedly reviewing unsafe or disturbing content is significant, and compensation should reflect that burden.''} 
P2 described a complementary approach of tailoring incentives to specific worker populations, including non-monetary rewards such as informational interviews or letters of recommendation.


\edit{At the same time, task designers worried that higher pay for RAI content work could itself become a source of pressure, drawing workers into tasks they might otherwise avoid.}
T9 described the problem: \textit{``all I'm saying is adding more money to those types of tasks would make people that are sensitive to some explicit or disturbing content to go ahead [and do them] without even looking back''} (T9).  
For these participants, improvements in data quality were a more compelling justification; T3 noted that they would be more open to higher pay if it came with some guarantee or strong expectation of better worker responses. To reduce this risk, T12 proposed that any pay increase should not be prominently publicly publicized, arguing that \textit{``it shouldn't be made public that they will be paid more for very severe tasks, because \dots [that] enhances the uncensored responses, where workers don't even pay attention to [the warning], since it is something they know pays more''} (T12). \edit{Overall, we surface a persistent tension around how workers articulate a desire to be paid more for the emotional and cognitive burdens of RAI content work, while task designers and platforms struggle to reconcile fair compensation with concerns about financial coercion and quality justification.}

\subsection{Post-Task Completion}
\edit{After tasks ended or workers chose to stop, there remained a final stage in which reflection, feedback, and redress may occur. Here, we found that workers sought both the option to share their experiences and concrete signs that task designers and platforms would respond when disclosure fell short. Task designers looked for feedback they could act on without overburdening, and platforms aimed to gather signals at scale while maintaining trust in the system. Our analysis focused on two related challenges: how to elicit high-quality feedback about risk without compromising worker agency, and how responsibility for harm was acknowledged or deflected when disclosure proved insufficient. Below, we discussed in each subsection how participants perceived risk disclosure (RQ1) and potential tensions (RQ2) among different stakeholder groups.}
\subsubsection{Encouraging quality feedback}
\edit{Participants across roles agreed that feedback from workers is crucial for evaluating risk disclosure, yet struggled to collect enough high-quality feedback in practice. Task designers and platform representatives described challenges in getting workers to respond, even to broad task questions not specifically focused on risk.}
P1 described difficulty collecting worker feedback at scale and called for more standardized feedback practices on their platform (P1). For task designers, feedback was especially important to understand whether risk disclosure was effective. T11 explained: 
\textit{``[until] you get feedback you don't know this exact content can be very, very disturbing. You might think that it's just moderate, but you don't know that it's very, very severe and extremely disturbing.''}

\edit{Workers, in turn, described feedback as one of the few mechanisms through which they could assert agency in the crowdsourcing process.}  W9 explained that after completing a task, \textit{``there's little I could do, but \dots after each task, you have a right [to be able to] leave a review. Leaving a review will make other other [workers] more aware of what was gonna [be in the task]''} (W9). Others saw feedback as an outlet for responding to harm when risk disclosure failed, including a space to complain about inadequate warnings. Taken together, these accounts demonstrate that feedback serves as both a diagnostic tool for task designers and a meaningful channel of communication for workers.

\edit{When discussing how to encourage more feedback, participants highlighted a tension between the desire for information and respect for worker agency.}
Several task designers reasoned that workers \textit{``should be able to choose things on their own free will''} (T13), but also that enforcing feedback may result in lower quality responses. Workers expressed a similar sentiment, emphasizing that they may be too tired or traumatized after a task to give feedback. W3 described such an instance: \textit{``I was so stressed that I just want[ed] to finish [the task] so I didn't give any feedback. [The task designer] asked for feedback. I just skipped it''} (W3). In this context, requiring feedback risks adding a burden at exactly the moment when workers may need to disengage the most.

\edit{Participants proposed compensation as one way to improve feedback quality, but this solution introduced its own tradeoffs.}
W1 suggested that paying for feedback could help workers \textit{``take [their] time providing it''} (W1). Some task designers, including T6 and T9, indicated they would be willing to pay for feedback, given its value in improving task design and risk disclosure. Others were more reluctant. T5 explained they would not try to get feedback if they needed to pay for it, and T12 felt that if feedback was \textit{``just three clicks,''} (T12) compensation would be unnecessary. Still, other participants worried that payment might incentivize disingenuous or \textit{``false feedback''} (T14). \edit{Overall, while design interventions such as compensated feedback hold promise, our findings surface persistent challenges in collecting feedback that is both ethically solicited and substantively informative, including how to model compensation and how to elicit nuanced responses without overburdening workers.}

\subsubsection{Mitigating harms from failed risk disclosure}
\edit{When risk was not adequately disclosed, workers often looked for acknowledgement or redress from task designers, but encountered a lack of clear accountability structures.}
Several workers felt it was important to receive a response from task designers when they raised problems with a risk disclosure process (e.g., W1 and W3). For some, responses from task designers helped validate the harm they experienced (e.g., W1 and W10). W8 shared that hearing back from a task designer  \textit{``makes me feel heard and sane at least. It tells me that you actually value my opinion''} (W8). 
For these participants, acknowledgment signaled that task designers recognized the harm and might adjust warnings for future workers.

Other workers struggled to imagine forms of redress beyond monetary compensation and worried about how such systems could be abused. W3 wanted the option to contact a task designer: \textit{``we should be able to contact their company''}; but struggled to envision what the task designer's organization could do other than provide compensation: \textit{``the other thing I'm thinking about is, if I contact the company, what are they going to do about it?''} (W3). When asked what they ideally wanted, W3 concluded \textit{``Oh compensate me, I guess''} (W3). At the same time, they acknowledged the risk that other workers might exploit task designers’ goodwill, underscoring a tension between recognizing harm and implementing compensation schemes that feel fair and sustainable.

\edit{Participants also pointed to well-being resources as a potential form of remediation, but disagreed on who should provide them and what responsibility that signaled.}
P1 suggested that platforms \textit{``offer resources. You can say like `if you experience issues, please reach out to [us]''} reasoning that if the platform is hosting such tasks \textit{``they should have a point of contact. The program manager, a mental health counselor, or something [else] in case people experience challenging issues in the work that they're doing''} (P1). P1 noted that their platform previously facilitated a red teaming task where they provided access to mental health professionals. Prior work similarly documents how some academic task designers within HCI-communities provide helplines or resource links as part of their protocols~\cite{qian2025locating, Zhang2024PerceptionsTriggerWarnings}. However, these practices remain discretionary, and workers were attuned to how responsibility was framed. W3 explained why task designers should offer their own resources:
\begin{quote}
    \textit{``It's just [task designers] fulfilling [some] righteousness because obviously nobody's gonna call the number \dots because you're putting a task for people to do and you're asking them to call [another organization]. Is it [that organization] that gives [workers] the task? Doesn't make sense to me. You give [workers] a task. You should be the one offering [workers] like a helpline and all that.''} - W3
\end{quote}
\edit{This reflection highlights the complexity of mitigation after failed disclosure. Workers interpret the design and placement of resources as signals of who is truly taking responsibility for harm; however, current practices diffuse accountability across individual designers, organizations, and platforms without clearly requiring any of them to respond.}

\section{Discussion}
\subsection{Design Implications}

\subsubsection{Front-load Worker Preparation Upon Platform Sign-up}
Our findings indicate that platforms should take a more active role in preparing workers for sensitive content tasks: sign-up surveys and opt-in checkboxes often function mainly as legal cover, even though workers saw platforms as responsible for enforcing protections and supporting well-being. \edit{This aligns with prior critiques of crowd platforms as powerful gatekeepers whose design choices distribute responsibility and risk across workers and requesters~\cite{irani2013turkopticon, fieseler_unfairness_2019}, but extends them by foregrounding how platforms explicitly mediate psychological harm in RAI content work.} Platforms could replace one-time sign-ups with structured opportunities for workers to set and revisit preferences, paired with brief onboarding modules and well-being strategies (for example, emotion regulation and cognitive reappraisal~\cite{spence2023content}) that introduce common sensitive content categories and prompt reflection on personal boundaries. Such preparation can foster self awareness, reduce unanticipated harm, and improve data quality, in line with calls for stronger preparation and resources for employed content moderators~\cite{qian2025aura, steiger_psychological_2021, steiger2022effects}, \edit{while our multi stakeholder findings locate this responsibility specifically at the platform sign up stage, where worker priorities for protection, task designer needs for a reliable labor pool, and platform concerns about liability first intersect.}


\subsubsection{Increase Specificity to Ensure Sustainable Worker Responses}
Workers, task designers, and platforms held mismatched assumptions about how warnings affected participation: designers feared that specificity would drive workers away, whereas workers repeatedly stressed that concrete details were what enabled informed choice. Our findings therefore point to prioritizing specific, content level disclosures over generic phrasing, echoing prior work showing that vague warnings on social media and in job ads leave both creators and readers uncertain about what risk is being managed~\cite{Zhang2024PerceptionsTriggerWarnings, bridgland2024meta} and that task designers often default to maximizing throughput and participation~\cite{ho2015incentivizing}. Greater specificity can support sustainability by shielding workers from unexpected harm and strengthening task designers reputations, which are key to participation and trust in crowdsourcing~\cite{gaikwad2016boomerang, irani2013turkopticon}. \edit{For task designers, more detailed warnings help align quality and throughput goals with workers’ need for informed consent, while for platforms they offer a clearer basis for enforcing participation rules and resolving disputes.} Future work could explore tools that help task designers author richer disclosures, such as AI driven keyword suggestions, severity taxonomies, or prompts informed by prior worker feedback. Reframing specificity as a benefit for both worker well-being and platform sustainability challenges the assumption that detail necessarily undermines participation.


\subsubsection{Encourage Feedback Through Facilitating Worker-Task Designer Relationship-Building}
Feedback was one of the few areas of broad agreement across stakeholders: workers treated it as a rare moment of agency, task designers as essential for improving warnings, and platforms as a way to maintain trust, yet it remained under-supported in most platform designs. Platforms could better surface and use feedback by integrating prompts into workflows, optionally compensating more substantive responses, and indicating when comments lead to concrete changes, while still allowing workers to decline after distressing tasks without penalty. Prior work has shown that relationships between workers and task designers shape trust and quality in crowdsourcing~\cite{irani2013turkopticon, qian2025locating}, but the role of feedback in mitigating risk disclosure failures remains underexplored; \edit{our findings add a more stakeholder-aware view in which workers use feedback to flag harm and warn peers, task designers rely on it to recalibrate disclosures, and platforms treat it as a signal for governance and policy enforcement. Designs that make these different purposes explicit can better align feedback channels such priorities.} Notably, early platform changes are already moving in this direction (see Appendix Figure \ref{fig:prolific-feedback} for Prolific's feedback option), but systematic evaluation is needed to generate empirical evidence that can better inform practice.

\footnotetext{Documented by an author completing a task on September, 2025}
\subsection{Negotiated Responsibility in Risk Disclosure}
Our study reveals that risk disclosure in AI data work is a socially negotiated process involving workers, task designers, and platforms. Consistent with prior research, we found that these groups hold distinct and at times conflicting assumptions about who is responsible for identifying, assessing, and communicating risks \cite{fieseler_unfairness_2019}. Workers assumed tasks were vetted for harm, while task designers assumed workers would self-select out of sensitive content. Platforms, meanwhile, frequently promote a vision of shared responsibility while retaining control over infrastructure and moderation policies. These mismatches lead to practical gaps in risk disclosure and reinforce structural ambiguities around accountability \cite{Suchman2002LocatedAccountabilities, widder_dislocated_2023}.

This ambiguity invites us to reframe the design challenge. Rather than only asking how risk should be disclosed, we need to ask who decides, when, and on what basis. These questions foreground power asymmetries in crowdsourcing infrastructures: workers have few ways to shape disclosure even as they bear the brunt of poorly disclosed risks; task designers may want to warn workers but are constrained by platform guidelines; and platforms provide tools and policies while remaining opaque about how risks are evaluated or escalated internally~\cite{roberts2019behind}. Framing risk disclosure as negotiated responsibility opens new avenues for design. Future tools might not only standardize formats, but also surface disagreements and support dialogue across stakeholder boundaries~\cite{fieseler_unfairness_2019}, for example, by showing discrepancies between worker and task-designer risk perceptions or allowing workers to annotate disclosures. This orientation draws on relational and distributed views of accountability~\cite{Suchman2002LocatedAccountabilities, CooperAccountability2022}, which treat disclosure not as a one-way transfer of information, but as an ongoing, situated process shaped by differing expertise, values, and institutional constraints.

\subsection{\edit{Beyond Risk Disclosure in Crowdsourced RAI Content Work}}

\edit{Our findings show that task-level risk disclosure was a necessary but insufficient response to the well-being risks in crowdsourced RAI content work. 
Up front, workers wanted clearer expectations about what they would see and tools to 
decide whether to engage with the work at all, aligning with previous literature 
~\cite{steiger_psychological_2021, steiger2022effects, dang_but_2018, das_fast_2020, qian2025aura}. For crowdsourced work conducted within the employer organization, the same organization often acts as employer, platform, and task designer, which gives it more direct leverage to standardize training and build such preparation into hiring and onboarding~\cite{blohm2018manage}. By contrast, crowd platforms distribute responsibilities across independent task designers and workers, which makes coordinated pre-task support harder to enforce. Moreover, unlike employed moderators, crowdworkers  lacked stable employment relationships or access to institutional care~\cite{irani2013turkopticon, silberman2018responsible}, so these practices would need to be adapted to short-term, geographically distributed workforces.}

\edit{After tasks, workers wanted ways to process and respond to unexpected harms, such as easier ways to pause, stop, and escalate concerns without penalty. Platform representatives and task designers, however, struggled to envision who would provide such support, how it would be funded, and how to manage it at scale, echoing tensions in internal content moderation where formal supports often exist but are overstretched or unevenly trusted~\cite{gray2019ghost, roberts2019behind}. For crowdsourcing work conducted outside of the employer organization, 
this gap between aspirational care and actual resources was even wider. Our results therefore suggest that risk disclosure mechanisms need to be coupled with broader commitments to care, such as partial payment policies when workers stop, and accessible support channels.}

\edit{These findings also raised a broader question: \textbf{Should risky or sensitive AI content work be crowdsourced at all, and if so, under what conditions?} Prior work has shown that microtasking can fragment judgment, obscure ethical complexity, and devalue the expertise needed for nuanced content decisions~\cite{berg2018digital}, concerns that are amplified in RAI development where aggregated annotations and red teaming outcomes shape downstream model behavior~\cite{diaz2022crowdworksheets, wang2022whose}. Task designers and platform representatives often justified crowdsourcing by pointing to diversity and scale~\cite{aroyo2023dices}, and some workers valued both the income and the chance to contribute, yet participants also acknowledged that current practices frequently extract insight from workers without commensurate protections or compensation. Rather than rejecting crowdsourcing outright, our results suggest reconfiguring how it is organized (e.g., shifting from isolated microtasks to longer form or collaborative workflows that provide more context, deliberation, and opportunities to opt out), and restructuring incentives so workers are not financially pressured to remain in harmful tasks (e.g., such as platform supported partial payment when workers stop for content related reasons). These directions resonate with trends in internal content moderation and red teaming that emphasize team based decision making, supervision, and clearer career pathways~\cite{qian2025aura, qian2025locating}, but adapting them to crowd platforms would require significant shifts in governance and business models, including a willingness to trade some efficiency and cost savings for care and deliberation.}

\subsection{Limitations and Future Work}
This study has limitations that should be addressed. Selection bias is possible across all three groups: participants who are more ethically attuned or institutionally empowered may have been more willing to engage in a study about risk and well-being. To mitigate this, we framed recruitment as a practical inquiry into improving RAI task design and platform process rather than an ethics study, and we observed variation in orientations within each group, including participants who reported minimal engagement with risk considerations. This range suggests our pool included different levels of reflexivity. While including platform representatives was a rare opportunity that broadened our understanding of risk disclosure challenges, these participants do not represent the full range of roles within platforms or the diversity of crowdsourcing platforms. Future work \edit{may elicit perspectives across roles and a wider range of crowdsourcing platforms}. Like many qualitative studies~\cite{tang2024ai}, our analysis relies on self reported accounts that may be shaped by social desirability pressures~\cite{bergen2020everything}, and our sample likely overrepresents people who have remained in RAI content work; we therefore encourage research that also includes those who have left the field.
\section{Conclusion}
The design space of risk disclosure in RAI content work is both essential and underexplored, especially as platforms increasingly route these tasks to crowdworkers who may be exposed to harmful content without robust consent processes. 
\edit{Our findings show that workers want clarity and control over what they see, task designers worry that detailed warnings will deter participation or hurt data quality, and platforms seek to protect workers while managing liability and meeting task designer demands.} Any intervention in this space must therefore grapple with these competing logics, rather than treating risk disclosure as a neutral or purely technical fix. We argue for systems that explicitly support negotiation and informed choice across stakeholders, as a step toward more sustainable and equitable models of risk disclosure in crowdsourced AI work.

\begin{acks}
This project was supported through Microsoft's AI and Society Fellowship\footnote{https://www.microsoft.com/en-us/research/academic-program/ai-society-fellows/}. Our work specifically focused on supporting the responsible AI red teaming human infrastructure. The first author was additionally supported by the NSF GRFP DGE2140739 and the National institute of Standards and Technology\footnote{ror.org/05xpvk416} under Federal Award ID Number 60NANB24D231 and Carnegie Mellon University\footnote{https://ror.org/05x2bcf33} AI Measurement Science and Engineering Center (AIMSEC). We note that we did not recruit from sponsoring organizations, and that our views are our own and do not represent those of our supporters or sponsors. Finally, our work would not have been possible without the bravery, care, and thoughtful engagement of our participants. 
\end{acks}

\bibliographystyle{ACM-Reference-Format}
\bibliography{_references, _references_aura, _references_crowdsourcing, _references_locating_risk}

@article{parker2019snowball,
  title={Snowball sampling},
  author={Parker, Charlie and Scott, Sam and Geddes, Alistair},
  journal={SAGE research methods foundations},
  year={2019},
  publisher={Sage}
}

@inproceedings{caine2016local,
  title={Local standards for sample size at CHI},
  author={Caine, Kelly},
  booktitle={Proceedings of the 2016 CHI conference on human factors in computing systems},
  pages={981--992},
  year={2016}
}

@article{blohm2018manage,
  title={How to manage crowdsourcing platforms effectively?},
  author={Blohm, Ivo and Zogaj, Shkodran and Bretschneider, Ulrich and Leimeister, Jan Marco},
  journal={California Management Review},
  volume={60},
  number={2},
  pages={122--149},
  year={2018},
  publisher={SAGE Publications Sage CA: Los Angeles, CA}
}

@article{bergen2020everything,
  title={“Everything is perfect, and we have no problems”: detecting and limiting social desirability bias in qualitative research},
  author={Bergen, Nicole and Labont{\'e}, Ronald},
  journal={Qualitative health research},
  volume={30},
  number={5},
  pages={783--792},
  year={2020},
  publisher={Sage Publications Sage CA: Los Angeles, CA}
}

@misc{StanfordHAI2025AIIndex,
  author       = {Stanford Institute for Human-Centered AI},
  title        = {Stanford HAI’s 2025 AI Index Reveals Record Growth in AI Capabilities, Investment, and Regulation},
  howpublished = {Press release},
  year         = {2025},
  url          = {https://www.businesswire.com/news/home/20250407539812/en/Stanford-HAIs-2025-AI-Index-Reveals-Record-Growth-in-AI-Capabilities-Investment-and-Regulation}
}

@article{jiang2023trade,
  title={A trade-off-centered framework of content moderation},
  author={Jiang, Jialun Aaron and Nie, Peipei and Brubaker, Jed R and Fiesler, Casey},
  journal={ACM Transactions on Computer-Human Interaction},
  volume={30},
  number={1},
  pages={1--34},
  year={2023},
  publisher={ACM New York, NY}
}

@article{jhaver2018online,
  title={Online harassment and content moderation: The case of blocklists},
  author={Jhaver, Shagun and Ghoshal, Sucheta and Bruckman, Amy and Gilbert, Eric},
  journal={ACM Transactions on Computer-Human Interaction (TOCHI)},
  volume={25},
  number={2},
  pages={1--33},
  year={2018},
  publisher={ACM New York, NY, USA}
}

@article{glossop2025co,
  title={Co-design of moderator training: Integrating knowledge from forum moderators, users and researchers with the improving peer online forums (iPOF) project},
  author={Glossop, Zoe and Jones, Steve and Ahmed, Saiqa and Caton, Neil and Collins, Gee and Haines, Jade and Jackson, Katherine and Lodge, Chris and Machin, Karen and Marshall, Paul and others},
  journal={Mental Health \& Prevention},
  volume={38},
  pages={200428},
  year={2025},
  publisher={Elsevier}
}

@misc{GrandViewResearch2024GenerativeAI,
  author       = {Grand View Research},
  title        = {Generative AI Market Size, Share \& Trends Analysis Report, 2025–2030},
  howpublished = {Market analysis summary},
  year         = {2024},
  url          = {https://www.grandviewresearch.com/industry-analysis/generative-ai-market-report}
}

@article{qian2025locating,
  title={Locating Risk: Task Designers and the Challenge of Risk Disclosure in RAI Content Work},
  author={Qian, Alice  and Shaw, Ryland and Dabbish, Laura and Suh, Jina and Shen, Hong},
  journal={arXiv preprint arXiv:2505.24246},
  year={2025}
}

@book{berastegui2021exposure,
  title={Exposure to psychosocial risk factors in the gig economy: a systematic review},
  author={B{\'e}rast{\'e}gui, Pierre},
  number={2021.01},
  year={2021},
  publisher={Report}
}

@online{ProlificResearcherSensitive2025,
  author       = {{Prolific Researcher Help Center}},
  title        = {How do I run a study with sensitive or disturbing content?},
  year         = {2025},
  month        = apr,
  note         = {Help Center article for researchers; provides guidance on conducting studies with sensitive or disturbing content and emphasises the importance of content warnings},
  url          = {https://researcher-help.prolific.com/en/article/ba774b},
  urldate      = {2025-09-05}
}

@online{ProlificAPIContentWarning2025,
  author       = {{Prolific API Help Center}},
  title        = {Adding a Content Warning to your Study},
  year         = {2025},
  month        = apr,
  note         = {API tutorial that explains how to assign “sensitive” or “explicit” content warnings via the Prolific API},
  url          = {https://api-help.prolific.com/en/article/51656c},
  urldate      = {2025-09-05}
}

@online{ProlificParticipantSensitive2025,
  author       = {{Prolific Participant Help Center}},
  title        = {Sensitive or disturbing study content},
  year         = {2025},
  month        = jun,
  note         = {Participant-facing guide that defines sensitive and disturbing content, advises how to interpret content warnings, and lists support resources},
  url          = {https://participant-help.prolific.com/en/article/6839b4},
  urldate      = {2025-09-05}
}

@online{MTurkTutorial2017,
  author  = {{Amazon Mechanical Turk}},
  title   = {Tutorial: Understanding Requirements and Qualifications},
  year    = {2017},
  month   = sep,
  note    = {Blog post on the Amazon Mechanical Turk Medium site explaining how requesters can define worker requirements and qualification tests},
  url     = {https://blog.mturk.com/tutorial-understanding-requirements-and-qualifications-99a26069fba2},
  urldate = {2025-09-05}
}

@article{gebrekidan2024content,
  title={Content moderation: The harrowing, traumatizing job that left many African data workers with mental health issues and drug dependency},
  author={Gebrekidan, Fasica B},
  journal={The Data Workers’ Inquiry. https://data-workers. org/fasica},
  year={2024}
}

@article{spence2025content,
  title={Content moderator mental health and associations with coping styles: replication and extension of previous studies},
  author={Spence, Ruth and DeMarco, Jeffrey},
  journal={Behavioral Sciences},
  volume={15},
  number={4},
  pages={487},
  year={2025},
  publisher={MDPI}
}

@article{martinez2024secondary,
  title={Secondary Trauma by Internet Content Moderation: a Case Report},
  author={Martinez-Sadurni, L and Casanovas, F and Llimona, C and Garcia, D and Rodriguez-Seoane, R and Castro, JI},
  journal={European Psychiatry},
  volume={67},
  number={Suppl 1},
  pages={S666},
  year={2024},
  publisher={Cambridge University Press}
}

@inproceedings{alemadi2024emotional,
  title={Emotional toll and coping strategies: Navigating the effects of annotating hate speech data},
  author={AlEmadi, Maryam M and Zaghouani, Wajdi},
  booktitle={Proceedings of the Workshop on Legal and Ethical Issues in Human Language Technologies@ LREC-COLING 2024},
  pages={66--72},
  year={2024}
}

@inproceedings{shashirekha2023trigger,
  title={Trigger Detection in Social Media Text},
  author={Shashirekha, Hosahalli Lakshmaiah and Hegde, Asha and Balouchzahi, Fazlourrahman},
  booktitle={Working Notes of CLEF 2023-Conference and Labs of the Evaluation Forum},
  year={2023}
}

@article{bridgland2024meta,
  title={A meta-analysis of the efficacy of trigger warnings, content warnings, and content notes},
  author={Bridgland, Victoria ME and Jones, Payton J and Bellet, Benjamin W},
  journal={Clinical Psychological Science},
  volume={12},
  number={4},
  pages={751--771},
  year={2024},
  publisher={Sage Publications Sage CA: Los Angeles, CA}
}

@article{bell2025warning,
  title={“Warning—This Content May Trigger Temporary Discomfort, Which Is Expected and Manageable”: The Effect of Modified Trigger-Warning Language on Reactions to Emotionally Provocative Content},
  author={Bell, Kathryn M and Howardson, Rebeka and Holmberg, Diane and Cornelius, Tara L},
  journal={Behavior Therapy},
  volume={56},
  number={2},
  pages={213--224},
  year={2025},
  publisher={Elsevier}
}

@article{vit2025use,
  title={The use of trigger warnings on social media: a text analysis study of X},
  author={Vit, Abigail Paradise and Puzis, Rami},
  journal={PloS one},
  volume={20},
  number={4},
  pages={e0322549},
  year={2025},
  publisher={Public Library of Science San Francisco, CA USA}
}

@article{charles2022typology,
  title={Typology of content warnings and trigger warnings: Systematic review},
  author={Charles, Ashleigh and Hare-Duke, Laurie and Nudds, Hannah and Franklin, Donna and Llewellyn-Beardsley, Joy and Rennick-Egglestone, Stefan and Gust, Onni and Ng, Fiona and Evans, Elizabeth and Knox, Emily and others},
  journal={PloS one},
  volume={17},
  number={5},
  pages={e0266722},
  year={2022},
  publisher={Public Library of Science San Francisco, CA USA}
}

@inproceedings{sharevski2022meaningful,
  title={Meaningful context, a red flag, or both? Preferences for enhanced misinformation warnings among US Twitter users},
  author={Sharevski, Filipo and Devine, Amy and Jachim, Peter and Pieroni, Emma},
  booktitle={Proceedings of the 2022 European Symposium on Usable Security},
  pages={189--201},
  year={2022}
}

@article{steen2013co,
  title={Co-design as a process of joint inquiry and imagination},
  author={Steen, Marc},
  journal={Design issues},
  volume={29},
  number={2},
  pages={16--28},
  year={2013},
  publisher={MIT Press One Rogers Street, Cambridge, MA 02142-1209, USA journals-info~…}
}

@inproceedings{ho2015incentivizing,
  title={Incentivizing high quality crowdwork},
  author={Ho, Chien-Ju and Slivkins, Aleksandrs and Suri, Siddharth and Vaughan, Jennifer Wortman},
  booktitle={Proceedings of the 24th International Conference on World Wide Web},
  pages={419--429},
  year={2015}
}

@article{rechkemmer2022understanding,
  title={Understanding the Microtask Crowdsourcing Experience for Workers with Disabilities: A Comparative View},
  author={Rechkemmer, Amy and Yin, Ming},
  journal={Proceedings of the ACM on Human-Computer Interaction},
  volume={6},
  number={CSCW2},
  pages={1--30},
  year={2022},
  publisher={ACM New York, NY, USA}
}

@inproceedings{varanasi2022feeling,
  title={Feeling proud, feeling embarrassed: Experiences of low-income women with crowd work},
  author={Varanasi, Rama Adithya and Siddarth, Divya and Seshadri, Vivek and Bali, Kalika and Vashistha, Aditya},
  booktitle={Proceedings of the 2022 CHI Conference on Human Factors in Computing Systems},
  pages={1--18},
  year={2022}
}

@article{liang2021embracing,
  title={Embracing four tensions in human-computer interaction research with marginalized people},
  author={Liang, Calvin A and Munson, Sean A and Kientz, Julie A},
  journal={ACM Transactions on Computer-Human Interaction (TOCHI)},
  volume={28},
  number={2},
  pages={1--47},
  year={2021},
  publisher={ACM New York, NY, USA}
}

@inproceedings{whiting2019fair,
  title={Fair work: Crowd work minimum wage with one line of code},
  author={Whiting, Mark E and Hugh, Grant and Bernstein, Michael S},
  booktitle={Proceedings of the AAAI Conference on Human Computation and Crowdsourcing},
  volume={7},
  pages={197--206},
  year={2019}
}

@inproceedings{gray2016crowd,
  title={The crowd is a collaborative network},
  author={Gray, Mary L and Suri, Siddharth and Ali, Syed Shoaib and Kulkarni, Deepti},
  booktitle={Proceedings of the 19th ACM conference on computer-supported cooperative work \& social computing},
  pages={134--147},
  year={2016}
}

@inproceedings{flores2020challenges,
  title={The challenges of crowd workers in rural and urban America},
  author={Flores-Saviaga, Claudia and Li, Yuwen and Hanrahan, Benjamin and Bigham, Jeffrey and Savage, Saiph},
  booktitle={Proceedings of the AAAI Conference on Human Computation and Crowdsourcing},
  volume={8},
  pages={159--162},
  year={2020}
}

@article{sutherland2018sharing,
  title={The sharing economy and digital platforms: A review and research agenda},
  author={Sutherland, Will and Jarrahi, Mohammad Hossein},
  journal={International Journal of Information Management},
  volume={43},
  pages={328--341},
  year={2018},
  publisher={Elsevier}
}

@article{huang2024design,
  title={Design Tensions in Online Freelancing Platforms: Using Speculative Participatory Design to Support Freelancers' Relationships with Clients},
  author={Huang, Jessica and Ma, Ning F and Rivera, Veronica A and Somani, Tabreek and Lee, Patrick Yung Kang and Mcgrenere, Joanna and Yoon, Dongwook},
  journal={Proceedings of the ACM on Human-Computer Interaction},
  volume={8},
  number={CSCW1},
  pages={1--28},
  year={2024},
  publisher={ACM New York, NY, USA}
}

@inproceedings{tang2024ai,
  title={AI failure cards: Understanding and supporting grassroots efforts to mitigate AI failures in homeless services},
  author={Tang, Ningjing and Zhi, Jiayin and Kuo, Tzu-Sheng and Kainaroi, Calla and Northup, Jeremy J and Holstein, Kenneth and Zhu, Haiyi and Heidari, Hoda and Shen, Hong},
  booktitle={Proceedings of the 2024 ACM Conference on Fairness, Accountability, and Transparency},
  pages={713--732},
  year={2024}
}

@inproceedings{vines2013configuring,
  title={Configuring participation: on how we involve people in design},
  author={Vines, John and Clarke, Rachel and Wright, Peter and McCarthy, John and Olivier, Patrick},
  booktitle={Proceedings of the SIGCHI conference on human factors in computing systems},
  pages={429--438},
  year={2013}
}

@inproceedings{mcinnis2016taking,
  title={Taking a HIT: Designing around rejection, mistrust, risk, and workers' experiences in Amazon Mechanical Turk},
  author={McInnis, Brian and Cosley, Dan and Nam, Chaebong and Leshed, Gilly},
  booktitle={Proceedings of the 2016 CHI conference on human factors in computing systems},
  pages={2271--2282},
  year={2016}
}

@inproceedings{saito2019turkscanner,
  title={Turkscanner: Predicting the hourly wage of microtasks},
  author={Saito, Susumu and Chiang, Chun-Wei and Savage, Saiph and Nakano, Teppei and Kobayashi, Tetsunori and Bigham, Jeffrey P},
  booktitle={The World Wide Web Conference},
  pages={3187--3193},
  year={2019}
}

@inproceedings{wu2017confusing,
  title={Confusing the crowd: Task instruction quality on amazon mechanical turk},
  author={Wu, Meng-Han and Quinn, Alexander},
  booktitle={Proceedings of the AAAI Conference on Human Computation and Crowdsourcing},
  volume={5},
  pages={206--215},
  year={2017}
}

@inproceedings{han2020crowd,
  title={Crowd worker strategies in relevance judgment tasks},
  author={Han, Lei and Maddalena, Eddy and Checco, Alessandro and Sarasua, Cristina and Gadiraju, Ujwal and Roitero, Kevin and Demartini, Gianluca},
  booktitle={Proceedings of the 13th international conference on web search and data mining},
  pages={241--249},
  year={2020}
}

@article{Suchman2002LocatedAccountabilities,
author = {Suchman, Lucy},
title = {Located accountabilities in technology production},
year = {2002},
issue_date = {September 2002},
publisher = {University of Aalborg},
address = {DNK},
volume = {14},
number = {2},
issn = {0905-0167},
journal = {Scand. J. Inf. Syst.},
month = sep,
pages = {91–105},
numpages = {15}
}

@misc{EU-DSA-2022,
  author       = {{European Parliament and Council of the European Union}},
  title        = {Regulation (EU) 2022/2065 on a Single Market for Digital Services (Digital Services Act)},
  howpublished = {Official Journal of the European Union, L 277, pp. 1–102},
  year         = {2022},
  note         = {Entered into force 16 November 2022}
}

@article{howe2006rise,
  title={The rise of crowdsourcing},
  author={Howe, Jeff and others},
  journal={Wired magazine},
  volume={14},
  number={6},
  pages={176--183},
  year={2006},
  publisher={S{\~a}o Francisco/CA}
}

@book{berg2018digital,
  title={Digital labour platforms and the future of work: Towards decent work in the online world},
  author={Berg, Janine and Furrer, Marianne and Harmon, Ellie and Rani, Uma and Silberman, M Six},
  year={2018},
  publisher={ILO}
}

@inproceedings{jhaver2022designing,
  title={Designing word filter tools for creator-led comment moderation},
  author={Jhaver, Shagun and Chen, Quan Ze and Knauss, Detlef and Zhang, Amy X},
  booktitle={Proceedings of the 2022 CHI conference on human factors in computing systems},
  pages={1--21},
  year={2022}
}

@article{jhaver2023personalizing,
  title={Personalizing content moderation on social media: User perspectives on moderation choices, interface design, and labor},
  author={Jhaver, Shagun and Qian, Alice and Chen, Quan Ze and Natarajan, Nikhila and Wang, Ruotong and Zhang, Amy X},
  journal={Proceedings of the ACM on Human-Computer Interaction},
  volume={7},
  number={CSCW2},
  pages={1--33},
  year={2023},
  publisher={ACM New York, NY, USA}
}

@article{gillespie2020expanding,
  title={Expanding the debate about content moderation: Scholarly research agendas for the coming policy debates},
  author={Gillespie, Tarleton and Aufderheide, Patricia and Carmi, Elinor and Gerrard, Ysabel and Gorwa, Robert and Matamoros-Fern{\'a}ndez, Ariadna and Roberts, Sarah T and Sinnreich, Aram and Myers West, Sarah},
  journal={Internet Policy Review},
  volume={9},
  number={4},
  pages={1--29},
  year={2020},
  publisher={Berlin: Alexander von Humboldt Institute for Internet and Society}
}

@misc{ohioknow,
  title = {On "I Know It When I See It"},
  author = {Ohio, Jacobellis v.},
  howpublished = {Legal case},
  year = {1964},
  note = {378 U.S. 184}
}

@article{dalal2024provocation,
  title={Provocation: Who benefits from" inclusion" in Generative AI?},
  author={Dalal, Samantha and Hall, Siobhan Mackenzie and Johnson, Nari},
  journal={arXiv preprint arXiv:2411.09102},
  year={2024}
}

@inproceedings{CooperAccountability2022,
author = {Cooper, A. Feder and Moss, Emanuel and Laufer, Benjamin and Nissenbaum, Helen},
title = {Accountability in an Algorithmic Society: Relationality, Responsibility, and Robustness in Machine Learning},
year = {2022},
isbn = {9781450393522},
publisher = {Association for Computing Machinery},
address = {New York, NY, USA},
url = {https://doi.org/10.1145/3531146.3533150},
doi = {10.1145/3531146.3533150},
booktitle = {Proceedings of the 2022 ACM Conference on Fairness, Accountability, and Transparency},
pages = {864–876},
numpages = {13},
location = {Seoul, Republic of Korea},
series = {FAccT '22}
}

@article{aroyo2023dices,
  title={Dices dataset: Diversity in conversational ai evaluation for safety},
  author={Aroyo, Lora and Taylor, Alex and Diaz, Mark and Homan, Christopher and Parrish, Alicia and Serapio-Garc{\'\i}a, Gregory and Prabhakaran, Vinodkumar and Wang, Ding},
  journal={Advances in Neural Information Processing Systems},
  volume={36},
  pages={53330--53342},
  year={2023}
}

@inproceedings{dillahunt2017reflections,
  title={Reflections on design methods for underserved communities},
  author={Dillahunt, Tawanna R and Erete, Sheena and Galusca, Roxana and Israni, Aarti and Nacu, Denise and Sengers, Phoebe},
  booktitle={Companion of the 2017 ACM conference on computer supported cooperative work and social computing},
  pages={409--413},
  year={2017}
}

@book{gray2019ghost,
  title={Ghost work: How to stop Silicon Valley from building a new global underclass},
  author={Gray, Mary L and Suri, Siddharth},
  year={2019},
  publisher={Harper Business}
}

@inproceedings{kuo2023understanding,
  title={Understanding frontline workers’ and unhoused individuals’ perspectives on ai used in homeless services},
  author={Kuo, Tzu-Sheng and Shen, Hong and Geum, Jisoo and Jones, Nev and Hong, Jason I and Zhu, Haiyi and Holstein, Kenneth},
  booktitle={Proceedings of the 2023 CHI Conference on Human Factors in Computing Systems},
  pages={1--17},
  year={2023}
}

@article{bodker_participatory_2018,
	title = {Participatory design that matters—Facing the big issues},
	volume = {25},
	issn = {15577325},
	doi = {10.1145/3152421},
	number = {1},
	journaltitle = {{ACM} Transactions on Computer-Human Interaction},
	author = {Bødker, Susanne and Kyng, Morten},
	date = {2018-02},
	note = {Publisher: Association for Computing Machinery},
}

@inbook{muller_participatory_nodate,
author = {Muller, Michael J.},
title = {Participatory design: the third space in HCI},
year = {2002},
isbn = {0805838384},
publisher = {L. Erlbaum Associates Inc.},
address = {USA},
booktitle = {The Human-Computer Interaction Handbook: Fundamentals, Evolving Technologies and Emerging Applications},
pages = {1051–1068},
numpages = {18}
}

@article{gutheim2012fantasktic,
  title={Fantasktic: Improving quality of results for novice crowdsourcing users},
  author={Gutheim, Philipp and Hartmann, Bj{\"o}rn},
  journal={EECS Dept., Univ. California, Berkeley, CA, USA, Tech. Rep. UCB/EECS-2012},
  volume={112},
  year={2012}
}

@inproceedings{finnerty2013keep,
  title={Keep it simple: Reward and task design in crowdsourcing},
  author={Finnerty, Ailbhe and Kucherbaev, Pavel and Tranquillini, Stefano and Convertino, Gregorio},
  booktitle={Proceedings of the Biannual Conference of the Italian Chapter of SIGCHI},
  pages={1--4},
  year={2013}
}

@inproceedings{gaikwad2016boomerang,
  author    = {Snehalkumar S. Gaikwad and Durim Morina and Adam Ginzberg and Catherine Mullings and Shirish Goyal and Dilrukshi Gamage and Michael S. Bernstein},
  title     = {Boomerang: Rebounding the Consequences of Reputation Feedback on Crowdsourcing Platforms},
  booktitle = {Proceedings of the 29th Annual Symposium on User Interface Software and Technology},
  pages     = {625--637},
  year      = {2016},
  publisher = {ACM},
  doi       = {10.1145/2984511.2984542},
  url       = {https://dl.acm.org/doi/10.1145/2984511.2984542}
}

@inproceedings{papoutsaki2015crowdsourcing,
  author    = {Alexandra Papoutsaki and Hua Guo and Danae Metaxa-Kakavouli and Connor Gramazio and Jeff Rasley and Wenting Xie and Guan Wang and Jeff Huang},
  title     = {Crowdsourcing from Scratch: A Pragmatic Experiment in Data Collection by Novice Requesters},
  booktitle = {Proceedings of the AAAI Conference on Human Computation and Crowdsourcing},
  volume    = {3},
  pages     = {140--149},
  year      = {2015},
  publisher = {AAAI Press},
  url       = {https://ojs.aaai.org/index.php/HCOMP/article/view/13230}
}

@article{xia2020privacy,
  author    = {Huichuan Xia and Brian McKernan},
  title     = {Privacy in Crowdsourcing: A Review of the Threats and Challenges},
  journal   = {Computer Supported Cooperative Work (CSCW)},
  volume    = {29},
  pages     = {263--301},
  year      = {2020},
  publisher = {Springer},
  doi       = {10.1007/s10606-020-09374-0},
  url       = {https://link.springer.com/article/10.1007/s10606-020-09374-0}
}

@inproceedings{xu2017incentivizing,
  author    = {Jia Xu and Hong Li and Yanhua Li and Dapeng Yang and Tingting Li},
  title     = {Incentivizing the Biased Requesters: Truthful Task Assignment Mechanisms in Crowdsourcing},
  booktitle = {2017 14th Annual IEEE International Conference on Sensing, Communication, and Networking (SECON)},
  pages     = {1--9},
  year      = {2017},
  publisher = {IEEE},
  doi       = {10.1109/SAHCN.2017.7964933},
  url       = {https://ieeexplore.ieee.org/document/7964933}
}

@inproceedings{kittur2008crowdsourcing,
  author    = {Aniket Kittur and Ed H. Chi and Bongwon Suh},
  title     = {Crowdsourcing User Studies with Mechanical Turk},
  booktitle = {Proceedings of the SIGCHI Conference on Human Factors in Computing Systems},
  pages     = {453--456},
  year      = {2008},
  publisher = {ACM},
  doi       = {10.1145/1357054.1357127},
  url       = {https://dl.acm.org/doi/10.1145/1357054.1357127}
}

@inproceedings{martin2014being,
  author    = {David Martin and Benjamin V. Hanrahan and Jacki O'Neill and Neha Gupta},
  title     = {Being a Turker},
  booktitle = {Proceedings of the 17th ACM Conference on Computer Supported Cooperative Work \& Social Computing},
  pages     = {224--235},
  year      = {2014},
  publisher = {ACM},
  doi       = {10.1145/2531602.2531663},
  url       = {https://dl.acm.org/doi/10.1145/2531602.2531663}
}

@inproceedings{salehi2015we,
  title={We are dynamo: Overcoming stalling and friction in collective action for crowd workers},
  author={Salehi, Niloufar and Irani, Lilly C and Bernstein, Michael S and Alkhatib, Ali and Ogbe, Eva and Milland, Kristy and Clickhappier},
  booktitle={Proceedings of the 33rd annual ACM conference on human factors in computing systems},
  pages={1621--1630},
  year={2015}
}

@article{salehi2018ink,
  title={Ink: Increasing worker agency to reduce friction in hiring crowd workers},
  author={Salehi, Niloufar and Bernstein, Michael S},
  journal={ACM Transactions on Computer-Human Interaction (TOCHI)},
  volume={25},
  number={2},
  pages={1--17},
  year={2018},
  publisher={ACM New York, NY, USA}
}

@article{silberman2018responsible,
  title={Responsible research with crowds: pay crowdworkers at least minimum wage},
  author={Silberman, M Six and Tomlinson, Bill and LaPlante, Rochelle and Ross, Joel and Irani, Lilly and Zaldivar, Andrew},
  journal={Communications of the ACM},
  volume={61},
  number={3},
  pages={39--41},
  year={2018},
  publisher={ACM New York, NY, USA}
}

@article{toxtli2021quantifying,
  title={Quantifying the invisible labor in crowd work},
  author={Toxtli, Carlos and Suri, Siddharth and Savage, Saiph},
  journal={Proceedings of the ACM on human-computer interaction},
  volume={5},
  number={CSCW2},
  pages={1--26},
  year={2021},
  publisher={ACM New York, NY, USA}
}

@article{zheng2011task,
  title={Task design, motivation, and participation in crowdsourcing contests},
  author={Zheng, Haichao and Li, Dahui and Hou, Wenhua},
  journal={International Journal of Electronic Commerce},
  volume={15},
  number={4},
  pages={57--88},
  year={2011},
  publisher={Taylor \& Francis}
}

@inproceedings{bragg2018sprout,
  title={Sprout: Crowd-powered task design for crowdsourcing},
  author={Bragg, Jonathan and Mausam and Weld, Daniel S},
  booktitle={Proceedings of the 31st annual acm symposium on user interface software and technology},
  pages={165--176},
  year={2018}
}

@article{allen2018design,
  title={Design crowdsourcing: The impact on new product performance of sourcing design solutions from the “crowd”},
  author={Allen, BJ and Chandrasekaran, Deepa and Basuroy, Suman},
  journal={Journal of Marketing},
  volume={82},
  number={2},
  pages={106--123},
  year={2018},
  publisher={SAGE Publications Sage CA: Los Angeles, CA}
}

@article{schlicher2021flexible,
  title={Flexible, self-determined… and unhealthy? An empirical study on somatic health among crowdworkers},
  author={Schlicher, Katharina D and Schulte, Julian and Reimann, Mareike and Maier, G{\"u}nter W},
  journal={Frontiers in Psychology},
  volume={12},
  pages={724966},
  year={2021},
  publisher={Frontiers Media SA}
}

@article{Spence2025ContentModeratorMentalHealth,
  author  = {Spence, Ruth and DeMarco, Jeffrey},
  title   = {Content Moderator Mental Health and Associations with Coping Styles: Replication and Extension of Previous Studies},
  journal = {Behavioural Sciences},
  volume  = {15},
  number  = {4},
  pages   = {487},
  year    = {2025},
  doi     = {10.3390/bs15040487},
  url     = {https://pmc.ncbi.nlm.nih.gov/articles/PMC12024403/}
}

@article{Zhang2024PerceptionsTriggerWarnings,
  author  = {Zhang, Xinyi and Gupta, Muskan and Altland, Emily and Lee, Sang Won},
  title   = {Understanding the Perceptions of Trigger Warning and Content Warning on Social Media Platforms in the U.S.},
  journal = {arXiv preprint arXiv:2504.15429},
  year    = {2024},
  url     = {https://arxiv.org/abs/2504.15429}
}

@incollection{smith1995semi,
  title={Semi-structured interviewing and qualitative analysis},
  author={Smith, Jonathan A},
  booktitle={Rethinking methods in psychology},
  pages={10--26},
  year={1995},
  publisher={SAGE publications Ltd}
}

@misc{prolific2025participant,
  title        = {Sensitive or Disturbing Study Content},
  author       = {{Prolific Participant Help Center}},
  year         = {2025},
  month        = {April},
  url          = {https://participant-help.prolific.com/en/article/6839b4},
  note         = {Last updated April 24, 2025}
}

@inproceedings{bharucha2023content,
  title={Content moderator startle response: a qualitative study},
  author={Bharucha, Timir and Steiger, Miriah E and Manchanda, Priyanka and Mere, Rainer and Huang, Xieyining},
  booktitle={International Congress on Information and Communication Technology},
  pages={217--232},
  year={2023},
  organization={Springer}
}

@article{qian2025aura,
  title={AURA: Amplifying Understanding, Resilience, and Awareness for Responsible AI Content Work},
  author={Qian, Alice and Amores, Judith and Shen, Hong and Czerwinski, Mary and Gray, Mary L and Suh, Jina},
  journal={Proceedings of the ACM on Human-Computer Interaction},
  volume={9}, 
  number={CSCW2}, 
  pages={1--45},
  year={2025}, 
  publisher={ACM New York, NY, USA}
}

@incollection{egelman2014crowdsourcing,
  title={Crowdsourcing in HCI research},
  author={Egelman, Serge and Chi, Ed H and Dow, Steven},
  booktitle={Ways of Knowing in HCI},
  pages={267--289},
  year={2014},
  publisher={Springer}
}

@article{clarke2017thematic,
  title={Thematic analysis},
  author={Clarke, Victoria and Braun, Virginia},
  journal={The journal of positive psychology},
  volume={12},
  number={3},
  pages={297--298},
  year={2017},
  publisher={Taylor \& Francis}
}

@inproceedings{diaz2022crowdworksheets,
  title={Crowdworksheets: Accounting for individual and collective identities underlying crowdsourced dataset annotation},
  author={D{\'\i}az, Mark and Kivlichan, Ian and Rosen, Rachel and Baker, Dylan and Amironesei, Razvan and Prabhakaran, Vinodkumar and Denton, Emily},
  booktitle={Proceedings of the 2022 ACM Conference on Fairness, Accountability, and Transparency},
  pages={2342--2351},
  year={2022}
}

@misc{Douek_2021, title={More content moderation is not always better}, url={https://www.wired.com/story/more-content-moderation-not-always-better/}, journal={Wired}, publisher={Conde Nast}, author={Douek, Evelyn}, year={2021}, month={Jun}}

@inproceedings{Michel2018ExContentMS,
  title={Ex-Content Moderator Sues Facebook, Saying Violent Images Caused Her PTSD - e-traces},
  author={Michel},
  year={2018},
  url={https://api.semanticscholar.org/CorpusID:150309904}
}

@article{arsht_2018_human,
  title={The human cost of online content moderation},
  author={Arsht, Andrew and Etcovitch, Daniel},
  journal={Harvard Journal of Law and Technology},
  volume={2},
  year={2018}
}

@article{pinchevski2023social,
  title={Social media’s canaries: content moderators between digital labor and mediated trauma},
  author={Pinchevski, Amit},
  journal={Media, Culture \& Society},
  volume={45},
  number={1},
  pages={212--221},
  year={2023},
  publisher={SAGE Publications Sage UK: London, England}
}

@article{spence2023content,
  title={Content moderators’ strategies for coping with the stress of moderating content online},
  author={Spence, Ruth and Harrison, Amy and Bradbury, Paula and Bleakley, Paul and Martellozzo, Elena and DeMarco, Jeffrey},
  journal={Journal of Online Trust and Safety},
  volume={1},
  number={5},
  year={2023}
}

@misc{Stackpole_2022, title={Content moderation is terrible by design}, url={https://hbr.org/2022/11/content-moderation-is-terrible-by-design#:~:text=Their%20work%20is%20largely%20invisible,or%20video%20at%20a%20time.}, journal={Harvard Business Review}, author={Stackpole, Thomas}, year={2022}, month={Nov}}

@article{roberts2016commercial,
  title={Commercial content moderation: Digital laborers' dirty work},
  author={Roberts, Sarah T},
  year={2016}
}

@misc{Dwoskin_2019, title={Inside facebook, the second-class workers who do the hardest job are waging a quiet battle}, url={https://www.washingtonpost.com/technology/2019/05/08/inside-facebook-second-class-workers-who-do-hardest-job-are-waging-quiet-battle/}, journal={The Washington Post}, publisher={WP Company}, author={Dwoskin, Elizabeth}, year={2019}, month={Jul}}

@book{roberts2019behind,
  title={Behind the screen},
  author={Roberts, Sarah T},
  year={2019},
  publisher={Yale University Press}
}

@inproceedings{dosono2019moderation,
  title={Moderation practices as emotional labor in sustaining online communities: The case of AAPI identity work on Reddit},
  author={Dosono, Bryan and Semaan, Bryan},
  booktitle={Proceedings of the 2019 CHI conference on human factors in computing systems},
  pages={1--13},
  year={2019}
}

@article{mann2025meta,
  title = {Leaked docs show how Meta's AI is trained to be safe, be 'flirty,' and navigate contentious topics},
  author = {Mann, Jyoti and Webb, Effie},
  journal = {Business Insider},
  date = {2025-05-06},
  url = {https://www.businessinsider.com/meta-ai-chatbot-training-scaleai-safe-flirty-leaked-documents-2025-5}
}

@misc{prolific2025sensitive,
  title        = {How do I run a study with sensitive or disturbing content?},
  author       = {{Prolific Researcher Help Center}},
  year         = {2025},
  month        = {April},
  url          = {https://researcher-help.prolific.com/en/article/ba774b},
  note         = {Last updated April 24, 2025}
}

@inproceedings{wang2022whose,
  title={Whose AI Dream? In search of the aspiration in data annotation.},
  author={Wang, Ding and Prabhat, Shantanu and Sambasivan, Nithya},
  booktitle={Proceedings of the 2022 CHI conference on human factors in computing systems},
  pages={1--16},
  year={2022}
}

@article{udupa2023ethical,
  title={Ethical scaling for content moderation: Extreme speech and the (in) significance of artificial intelligence},
  author={Udupa, Sahana and Maronikolakis, Antonis and Wisiorek, Axel},
  journal={Big Data \& Society},
  volume={10},
  number={1},
  pages={20539517231172424},
  year={2023},
  publisher={SAGE Publications Sage UK: London, England}
}

@article{gillespie2010politics,
  title={The politics of ‘platforms’},
  author={Gillespie, Tarleton},
  journal={New media \& society},
  volume={12},
  number={3},
  pages={347--364},
  year={2010},
  publisher={Sage Publications Sage UK: London, England}
}

@inproceedings{steiger2022effects,
  title={Effects of a Novel Resiliency Training Program for Social Media Content Moderators},
  author={Steiger, Miriah and Bharucha, Timir J and Torralba, Wilfredo and Savio, Marlyn and Manchanda, Priyanka and Lutz-Guevara, Rachel},
  booktitle={Proceedings of Seventh International Congress on Information and Communication Technology: ICICT 2022, London, Volume 4},
  pages={283--298},
  year={2022},
  organization={Springer}
}

@inproceedings{steiger_psychological_2021,
	location = {Yokohama Japan},
	title = {The Psychological Well-Being of Content Moderators: The Emotional Labor of Commercial Moderation and Avenues for Improving Support},
	isbn = {978-1-4503-8096-6},
	url = {https://dl.acm.org/doi/10.1145/3411764.3445092},
	doi = {10.1145/3411764.3445092},
	shorttitle = {The Psychological Well-Being of Content Moderators},
	eventtitle = {{CHI} '21: {CHI} Conference on Human Factors in Computing Systems},
	pages = {1--14},
	booktitle = {Proceedings of the 2021 {CHI} Conference on Human Factors in Computing Systems},
	publisher = {{ACM}},
	author = {Steiger, Miriah and Bharucha, Timir J and Venkatagiri, Sukrit and Riedl, Martin J. and Lease, Matthew},
	urldate = {2023-04-07},
	date = {2021-05-06},
	langid = {english},
}

@article{schopke-gonzalez_why_2022,
	title = {Why do volunteer content moderators quit? Burnout, conflict, and harmful behaviors},
	issn = {1461-4448, 1461-7315},
	url = {http://journals.sagepub.com/doi/10.1177/14614448221138529},
	doi = {10.1177/14614448221138529},
	shorttitle = {Why do volunteer content moderators quit?},
	pages = {146144482211385},
	journaltitle = {New Media \& Society},
	shortjournal = {New Media \& Society},
	author = {Schöpke-Gonzalez, Angela M. and Atreja, Shubham and Shin, Han Na and Ahmed, Najmin and Hemphill, Libby},
	urldate = {2023-05-30},
	date = {2022-12-04},
	langid = {english},
}

@article{ruckenstein_re-humanizing_2020,
	title = {Re-humanizing the platform: Content moderators and the logic of care},
	volume = {22},
	pages = {1026--1042},
	number = {6},
	journaltitle = {New media \& society},
	author = {Ruckenstein, Minna and Turunen, Linda Lisa Maria},
	year = {2020},
	note = {Publisher: Sage Publications Sage {UK}: London, England},
}

@article{dang_but_2018,
	title = {But who protects the moderators? the case of crowdsourced image moderation},
	journaltitle = {{arXiv} preprint {arXiv}:1804.10999},
	author = {Dang, Brandon and Riedl, Martin J and Lease, Matthew},
	year = {2018},
}

@inproceedings{das_fast_2020,
	title = {Fast, accurate, and healthier: Interactive blurring helps moderators reduce exposure to harmful content},
	volume = {8},
	pages = {33--42},
	booktitle = {Proceedings of the {AAAI} Conference on Human Computation and Crowdsourcing},
	author = {Das, Anubrata and Dang, Brandon and Lease, Matthew},
	year = {2020},
}

@misc{newton_trauma_2019,
	title = {The trauma floor},
	url = {https://www.theverge.com/2019/2/25/18229714/cognizant-facebook-content-moderator-interviews-trauma-working-conditions-arizona},
	publisher = {The Verge},
	author = {Newton, Casey},
	date = {2019-02},
	note = {Publication Title: The Verge},
}

@inproceedings{irani2013turkopticon,
  title={Turkopticon: Interrupting worker invisibility in amazon mechanical turk},
  author={Irani, Lilly C and Silberman, M Six},
  booktitle={Proceedings of the SIGCHI conference on human factors in computing systems},
  pages={611--620},
  year={2013}
}

@article{fieseler_unfairness_2019,
	title = {Unfairness by {Design}? {The} {Perceived} {Fairness} of {Digital} {Labor} on {Crowdworking} {Platforms}},
	volume = {156},
	issn = {0167-4544, 1573-0697},
	shorttitle = {Unfairness by {Design}?},
	url = {http://link.springer.com/10.1007/s10551-017-3607-2},
	doi = {10.1007/s10551-017-3607-2},
	language = {en},
	number = {4},
	urldate = {2025-04-21},
	journal = {Journal of Business Ethics},
	author = {Fieseler, Christian and Bucher, Eliane and Hoffmann, Christian Pieter},
	month = jun,
	year = {2019},
	pages = {987--1005},
}

@inproceedings{hsieh_designing_2023,
	address = {New York, NY, USA},
	series = {{CHIWORK} '23},
	title = {Designing {Individualized} {Policy} and {Technology} {Interventions} to {Improve} {Gig} {Work} {Conditions}},
	isbn = {9798400708077},
	url = {https://dl.acm.org/doi/10.1145/3596671.3598576},
	doi = {10.1145/3596671.3598576},
	urldate = {2025-05-05},
	booktitle = {Proceedings of the 2nd {Annual} {Meeting} of the {Symposium} on {Human}-{Computer} {Interaction} for {Work}},
	publisher = {Association for Computing Machinery},
	author = {Hsieh, Jane and Adisa, Oluwatobi and Bafna, Sachi and Zhu, Haiyi},
	month = sep,
	year = {2023},
	pages = {1--9},
}

@article{widder_dislocated_2023,
	title = {Dislocated {Accountabilities} in the {AI} {Supply} {Chain}: {Modularity} and {Developers}' {Notions} of {Responsibility}},
	volume = {10},
	issn = {2053-9517, 2053-9517},
	shorttitle = {Dislocated {Accountabilities} in the {AI} {Supply} {Chain}},
	url = {http://arxiv.org/abs/2209.09780},
	doi = {10.1177/20539517231177620},
	number = {1},
	urldate = {2025-05-06},
	journal = {Big Data \& Society},
	author = {Widder, David Gray and Nafus, Dawn},
	month = jan,
	year = {2023},
	note = {arXiv:2209.09780 [cs]},
	pages = {20539517231177620},
}
\section{Appendix}

\subsection{Examples of Co-Design Probes}


\begin{figure}[H]
    \centering
    \begin{subfigure}[b]{0.48\textwidth}
        \centering
        \includegraphics[width=\textwidth]{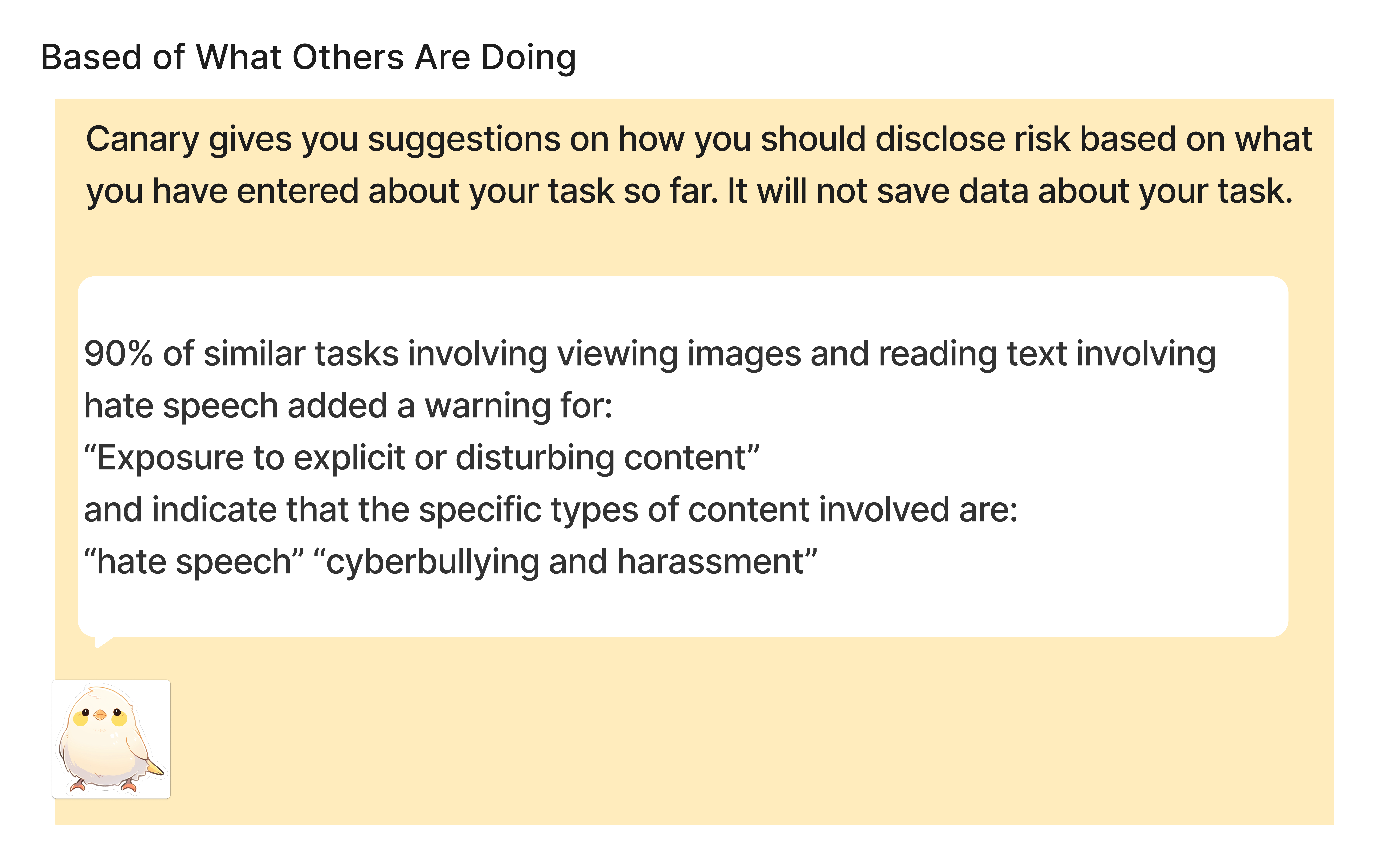}
        \caption{Suggestion based on what other task designers have done}
        \label{fig:AI-based-on-others}
    \end{subfigure}
    \hfill
    \begin{subfigure}[b]{0.48\textwidth}
        \centering
        \includegraphics[width=\textwidth]{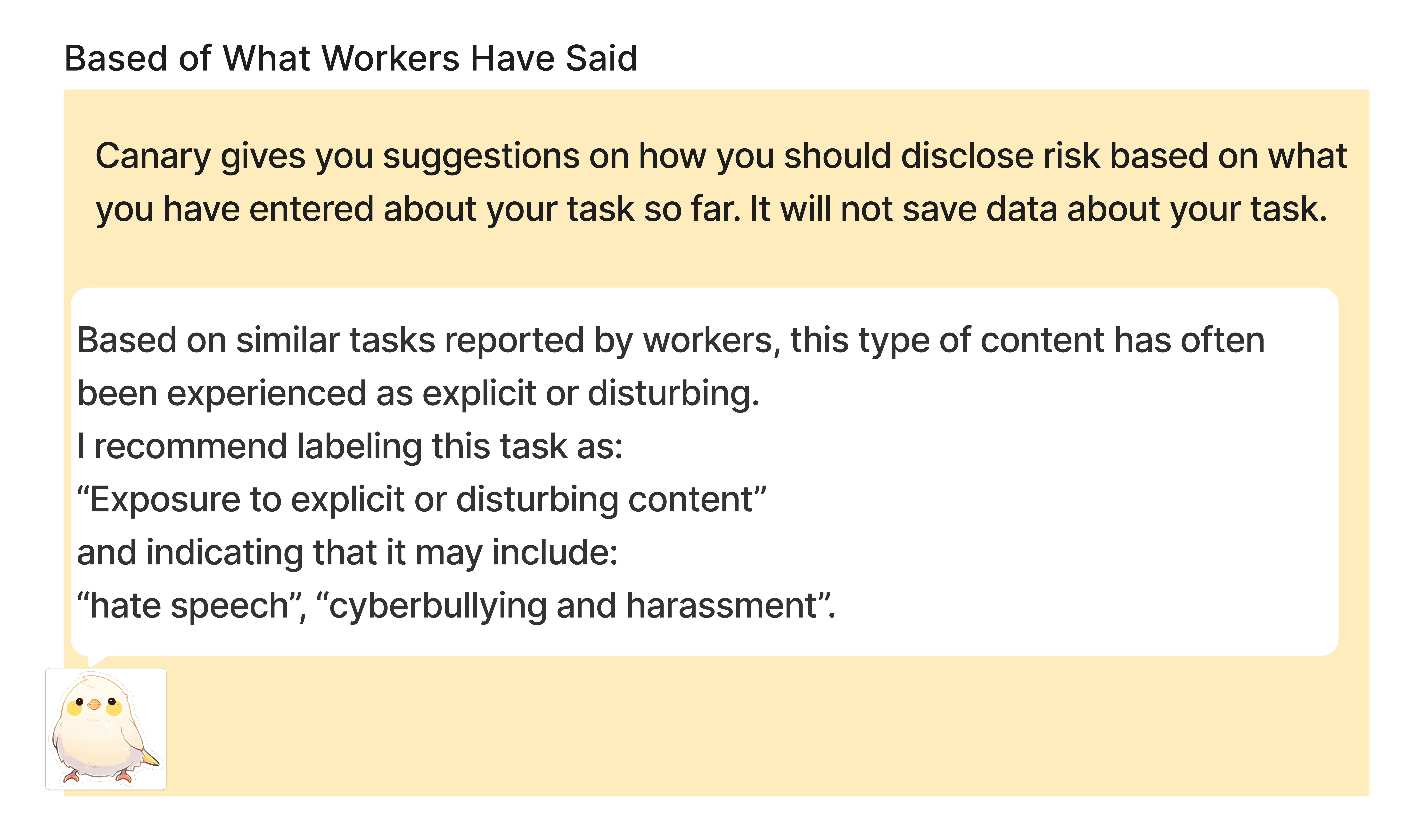} 
        \caption{Suggestion based on what workers have said}
        \label{fig:AI-based-on-workers}
    \end{subfigure}
    \caption{AI warning suggestions (Part 2)}
    \label{fig:task-designer-AI-2}
\end{figure}

\twocolumn[
\subsection{Additional participant demographics}\label{appendix:demographics}]
\begin{table}
\centering
\small
\begin{tabular}{p{2cm} p{5cm}}
\toprule
\textbf{Category} & \textbf{Summary} \\
\midrule
Organization type 
& Academia (4), Industry (11)\\
Organization size (number of employees)
& 10-40 (1), 50-249 (2), 1,000-4,999 (4), 5,000-24,999 (6), NA (1)\\

Task types 
& Data generation (8), Data labeling (11), Content removal (10), Adversarial prompting (6),  \\

Types of sensitive content 
& Misinformation (2), Bias\textbackslash Stereotyping (5), Hatespeech (6), Harassment and Bullying (5), Violent and graphic content (5), Terrorism and extremism (3), Self-harm and suicide (3)\\

Typical target responses per task
& < 100 responses (2), < 100 (2), < 1,000 (2), < 10,000 (7), < 100,000 (1), NA (2)\\

Platforms used 
& Prolific (10), Amazon Mechanical Turk (7), Remotasks (2), TaskUs (2) \\
\bottomrule
\end{tabular}
\caption{Summary of task designer participant characteristics across organization size, task types, sensitive content handled, target number of responses per task, and platforms used. Note that aggregate counts are not mutually exclusive, as some participants used more than one platform, for example. NA counts represent areas where participants chose not to disclose additional details.}
\label{tab:participant_summary_task_designer}
\end{table}

\begin{table}
\centering
\small
\begin{tabular}{p{2cm} p{5cm}}
\toprule
\textbf{Category} & \textbf{Summary} \\
\midrule
Task types 
& Data generation (9), Data labeling (11), Content removal (10), Adversarial prompting (7), \\

Types of sensitive content 
& Misinformation (6), Bias\textbackslash Stereotyping (6), Hatespeech (6), Harassment and Bullying (5), Violent and graphic content (6), Terrorism and extremism (1), Sexually explicit content (5), Self-harm and suicide (3), Illegal activities (2), Scams and fraud (2) \\

Typical number of tasks completed per day
& 1-10 (3), 11-50 (8), 51-100 (1) \\

Platforms used 
&  Prolific (8), Amazon Mechanical Turk (8), DataAnnotation (2), TaskUs (2)\\
\bottomrule
\end{tabular}
\caption{Summary of worker participant characteristics across task types, sensitive content handled, target number of responses per task, and platforms used. Note that aggregate counts are not mutually exclusive, as some participants used more than one platform, for example.}
\label{tab:participant_summary_worker}
\end{table}

\end{document}